\documentclass[
   aps,
   prb,
   twocolumn,
   reprint,
   superscriptaddress,
   floatfix,
   citeautoscript,
   longbibliography,
]{revtex4-2}
\usepackage[T1]{fontenc} 
\usepackage[utf8]{inputenc}
\usepackage{newtxtext}
\usepackage{newtxmath}
\usepackage{graphicx}
\usepackage{siunitx}
\usepackage{chemformula}
\usepackage{booktabs}
\usepackage[english]{babel}
\usepackage{csquotes}
\usepackage{microtype}
\usepackage[
   unicode,
   colorlinks = true,
   allcolors = blue
]{hyperref}
\usepackage{cleveref}
\usepackage{mathrsfs}

\usepackage{glossaries}
\glsdisablehyper

\newacronym{1d}{1D}{one-dimensional}
\newacronym{2d}{2D}{two-dimensional}
\newacronym{3d}{3D}{three-dimensional}

\newacronym{ac}{AC}{alternating current}
\newacronym{afm}{AFM}{atomic force microscopy}
\newacronym{alc}{ALC}{avoided level crossing}
\newacronym{api}{API}{application programming interface}
\newacronym{ariel}{ARIEL}{Advanced Rare Isotope Laboratory}
\newacronym{arpes}{ARPES}{angle-resolved photoemission spectroscopy}

\newacronym[sort={b-NMR}]{bnmr}{\ensuremath{\beta}-NMR}{\ensuremath{\beta}-detected nuclear magnetic resonance}
\newacronym[sort={b-NQR}]{bnqr}{\ensuremath{\beta}-NQR}{\ensuremath{\beta}-detected nuclear quadrupole resonance}
\newacronym{bca}{BCA}{binary collision approximation}
\newacronym{bpp}{BPP}{Bloembergen-Purcell-Pound}
\newacronym{bsc}{BSC}{\ch{Bi2Se3:Ca}}
\newacronym{btm}{BTM}{\ch{Bi2Te3:Mn}}
\newacronym{bts}{BTS}{\ch{Bi2Te2Se}}

\newacronym{camp}{CAMP}{control and monitor program}
\newacronym{ccd}{CCD}{charge-coupled device}
\newacronym{cdw}{CDW}{charge density wave}
\newacronym{cgs}{CGS}{centimetre-gram-second system of units}
\newacronym{cmms}{CMMS}{Centre for Molecular and Materials Science}
\newacronym{codata}{CODATA}{Committee on Data for Science and Technology}
\newacronym{cpu}{CPU}{central processing unit}
\newacronym{create}{CREATE}{Collaborative Research and Training Experience Program}
\newacronym{cw}{CW}{continuous wave}

\newacronym{daq}{DAQ}{data acquisition}
\newacronym{dc}{DC}{direct current}
\newacronym{dft}{DFT}{density functional theory}
\newacronym{dos}{DOS}{density of states}
\newacronym{dqt}{DQT}{double-quantum transition}

\newacronym{efg}{EFG}{electric field gradient}
\newacronym{epr}{EPR}{electron paramagnetic resonance}
\newacronym{esr}{EPR}{electron spin resonance}
\newacronym{endor}{ENDOR}{electron nuclear double resonance}
\newacronym{epics}{EPICS}{Experimental Physics and Industrial Control System}

\newacronym{fft}{FFT}{fast Fourier transform}
\newacronym{fom}{FoM}{figure of merit}
\newacronym{fwhm}{FWHM}{full width at half maximum}

\newacronym{gga}{GGA}{generalized gradient approximation}

\newacronym{hv}{HV}{high-voltage}
\newacronym{hwhm}{HWHM}{half width at half maximum}

\newacronym{is}{IS}{impedance spectroscopy}
\newacronym{isac}{ISAC}{isotope separator and accelerator}
\newacronym{isol}{ISOL}{isotope separation online}
\newacronym{isosim}{IsoSiM}{Isotopes for Science and Medicine}

\newacronym{lcao}{LCAO}{linear combination of atomic orbitals}
\newacronym{lda}{LDA}{local density approximation}
\newacronym{leis}{LEIS}{low-energy ion scattering}
\newacronym{lib}{LIB}{lithium-ion battery}
\newacronym{lsat}{LSAT}{\ch{(La,Sr)(Al,Ta)O3}}

\newacronym{mas}{MAS}{magic angle spinning}
\newacronym{mpms}{MPMS}{magnetic property measurement system}
\newacronym{mbe}{MBE}{molecular beam epitaxy}
\newacronym{md}{MD}{molecular dynamics}
\newacronym{midas}{MIDAS}{Maximum Integrated Data Acquisition System}
\newacronym{mit}{MIT}{metal-insulator transition}
\newacronym{mnr}{MNR}{Meyer-Neldel rule}
\newacronym{mqt}{mqt}{multi-quantum transition}
\newacronym{mud}{MUD}{muon data}

\newacronym{nbm}{NBM}{neutral beam monitor}
\newacronym{neb}{NEB}{nudged elastic band}
\newacronym{nim}{NIM}{nuclear instrumentation module}
\newacronym{nmr}{NMR}{nuclear magnetic resonance}
\newacronym{no}{NO}{nuclear orientation}
\newacronym{nqr}{NQR}{nuclear quadrupole resonance}
\newacronym{nserc}{NSERC}{Natural Sciences and Engineering Research Council of Canada}

\newacronym{oa}{OA}{optical absorption}

\newacronym{pac}{PAC}{perturbed angular correlation}
\newacronym{pad}{PAD}{perturbed angular distribution}
\newacronym{pas}{PAS}{principle axis system}
\newacronym{pld}{PLD}{pulsed laser deposition}
\newacronym{ppms}{PPMS}{physical property measurement system}

\newacronym{qens}{QENS}{quasielastic neutron scattering}
\newacronym{ql}{QL}{quintuple layer}
\newacronym{qo}{QO}{quantum oscillations}

\newacronym{rbs}{RBS}{Rutherford backscattering}
\newacronym{rf}{RF}{radio frequency}
\newacronym{rheed}{RHEED}{reflection high-energy electron diffraction}
\newacronym{rib}{RIB}{radioactive ion beam}
\newacronym{rkky}{RKKY}{Ruderman–Kittel–Kasuya–Yosida}

\newacronym{sae}{SAE}{spin-alignment echo}
\newacronym{si}{SI}{International System of Units}
\newacronym{sims}{SIMS}{secondary ion mass spectrometry}
\newacronym{slr}{SLR}{spin-lattice relaxation}
\newacronym[sort={S/N}]{snr}{\textit{S}/\textit{N}}{signal-to-noise ratio}
\newacronym{squid}{SQUID}{superconducting quantum interference device}
\newacronym{srim}{SRIM}{Stopping and Range of Ions in Matter}
\newacronym{ssid}{SSID}{solid-state ionic device}
\newacronym{ssr}{SSR}{spin-spin relaxation}
\newacronym{stm}{STM}{scanning tunnelling microscopy}
\newacronym{sts}{STS}{scanning tunnelling spectroscopy}

\newacronym{ti}{TI}{topological insulator}
\newacronym{trim}{TRIM}{Transport and Range of Ions in Matter}
\newacronym{tss}{TSS}{topological surface state}
\newacronym{tmd}{TMD}{transition metal dichalcogenide}

\newacronym{uhv}{UHV}{ultra-high vacuum}

\newacronym{vdw}{vdW}{van der Waals}

\newacronym{xrd}{XRD}{x-ray diffraction}
\newacronym{xrr}{XRR}{x-ray reflection}

\newacronym{ybco}{YBCO}{\ch{YBa2Cu3O_{6+x}}}
\newacronym{ysz}{YSZ}{yttria-stabilized zirconia}

\newacronym[sort={muSR}]{musr}{\ensuremath{\mu}SR}{muon spin rotation/relaxation/resonance}
\newacronym{alc-musr}{ALC-μSR}{avoided level crossing muon spin rotation}
\newacronym{le-musr}{LE-μSR}{low energy muon spin rotation}
\newacronym{lf-musr}{LF-μSR}{longitudinal field muon spin rotation}
\newacronym{rf-musr}{RF-μSR}{radio frequency muon spin rotation}
\newacronym{tf-musr}{TF-μSR}{transverse field muon spin rotation}
\newacronym{zf-musr}{ZF-μSR}{zero field muon spin rotation}

\newcommand{\latin}[1]{#1}

\DeclareSIUnit\emu{emu}
\DeclareSIUnit\gauss{G}
\DeclareSIUnit\ppm{ppm}

\begin{document} 

\title{Local electronic and magnetic properties of the doped topological insulators \ch{Bi2Se3:Ca} and \ch{Bi2Te3:Mn} investigated using ion-implanted \ch{^{8}Li} $\beta$-NMR}

\newcommand{\ubcchem}{Department of Chemistry, University of British Columbia, Vancouver, BC V6T~1Z1, Canada}
\newcommand{\ubcsbqmi}{Stewart Blusson Quantum Matter Institute, University of British Columbia, Vancouver, BC V6T~1Z4, Canada}
\newcommand{\ubcphas}{Department of Physics and Astronomy, University of British Columbia, Vancouver, BC V6T~1Z1, Canada}
\newcommand{\triumf}{TRIUMF, 4004 Wesbrook Mall, Vancouver, BC V6T~2A3, Canada}
\newcommand{\sfuchem}{Department of Chemistry, Simon Fraser University, Burnaby, BC V5A~1S6, Canada}
\newcommand{\princetonchem}{Department of Chemistry, Princeton University, Princeton, New Jersey 08544, USA}
\newcommand{\mstphys}{Department of Physics, Missouri University of Science and Technology, Rolla, Missouri 65409, USA}

\author{Ryan~M.~L.~McFadden}
\email{rmlm@chem.ubc.ca}
\affiliation{\ubcchem}
\affiliation{\ubcsbqmi}

\author{Aris~Chatzichristos}
\affiliation{\ubcsbqmi}
\affiliation{\ubcphas}

\author{David~L.~Cortie}
\altaffiliation{Current address: Institute for Superconducting and Electronic Materials, Australian Institute for Innovative Materials, University of Wollongong, North Wollongong, NSW 2500, Australia}
\affiliation{\ubcchem}
\affiliation{\ubcsbqmi}
\affiliation{\ubcphas}

\author{Derek~Fujimoto}
\affiliation{\ubcsbqmi}
\affiliation{\ubcphas}

\author{Yew~San~Hor}
\affiliation{\mstphys}

\author{Huiwen~Ji}
\altaffiliation{Current address: Department of Materials Science and Engineering, University of California, Berkeley, Berkeley, CA 94720, USA}
\affiliation{\princetonchem}

\author{Victoria~L.~Karner}
\affiliation{\ubcchem}
\affiliation{\ubcsbqmi}

\author{Robert~F.~Kiefl}
\affiliation{\ubcsbqmi}
\affiliation{\ubcphas}
\affiliation{\triumf}

\author{C.~D.~Philip~Levy}
\affiliation{\triumf}

\author{Ruohong~Li}
\affiliation{\triumf}

\author{Iain~McKenzie}
\affiliation{\triumf}
\affiliation{\sfuchem}

\author{Gerald~D.~Morris}
\affiliation{\triumf}

\author{Matthew~R.~Pearson}
\affiliation{\triumf}

\author{Monika~Stachura}
\affiliation{\triumf}

\author{Robert~J.~Cava}
\affiliation{\princetonchem}

\author{W.~Andrew~MacFarlane}
\email{wam@chem.ubc.ca}
\affiliation{\ubcchem}
\affiliation{\ubcsbqmi}
\affiliation{\triumf}

\date{\today}

\begin{abstract}
We report \acrshort{bnmr} measurements in \acrlong{bsc} and \acrlong{btm} single crystals using \ch{^{8}Li^{+}} implanted to depths on the order of \SI{100}{\nano\meter}.
Above \SI{\sim 200}{\kelvin}, \acrlong{slr} reveals diffusion of \ch{^{8}Li^{+}}, with activation energies of \SI{\sim 0.4}{\electronvolt} (\SI{\sim 0.2}{\electronvolt}) in \acrlong{bsc} (\acrlong{btm}).
At lower temperatures, the \acrshort{nmr} properties are those of a heavily doped semiconductor in the metallic limit, with Korringa relaxation and a small, negative, temperature-dependent Knight shift in \acrlong{bsc}.
From this, we make a detailed comparison with the isostructural tetradymite \acrlong{bts} [McFadden \latin{et al.}, \href{https://doi.org/10.1103/PhysRevB.99.125201}{Phys Rev.\ B \textbf{99}, 125201 (2019)}].
In the magnetic \acrlong{btm}, the effects of the dilute \ch{Mn} moments predominate, but remarkably the \ch{^{8}Li} signal is not wiped out through the magnetic transition at \SI{13}{\kelvin}, with a prominent critical peak in the \acrlong{slr} that is suppressed in a high applied field.
This detailed characterization of the \ch{^{8}Li} \acrshort{nmr} response is an important step towards using depth-resolved \acrshort{bnmr} to study the low-energy properties of the chiral \acrlong{tss} in the \ch{Bi2Ch3} tetradymite \acrlongpl{ti}.
\end{abstract}

\maketitle
\glsresetall

\section{Introduction \label{sec:introduction}}

The bismuth chalcogenides \ch{Bi2Ch3} (\ch{Ch} = \ch{S}, \ch{Se}, or \ch{Te}) of the layered tetradymite structure are an interesting class of highly two dimensional narrow bandgap semiconductors.
Strong spin-orbit coupling inverts the energy ordering of their bands, making them bulk \glspl{ti} characterized by a single Dirac cone at the Brillouin zone centre and a topologically protected metallic surface state~\cite{2009-Hsieh-N-460-1101, 2009-Zhang-NP-5-438}.
This has augmented longstanding interest in their thermoelectric properties with significant efforts (theoretical and experimental) to understand their electronic properties in detail~\cite{2013-Cava-JMCC-1-3176, 2017-Heremans-NRM-2-17049}.
Consisting of weakly interacting \ch{Ch-Bi-Ch-Bi-Ch} atomic \glspl{ql} (see \latin{e.g.}, Figure~1 in Ref.~\onlinecite{2019-McFadden-PRB-99-125201}), they can also accommodate intercalant species such as \ch{Li^{+}} in the \gls{vdw} gap between \glspl{ql}~\cite{1988-Paraskevopoulos-MSEB-1-147, 1989-Julien-SSI-36-113, 2010-Bludska-JSSC-183-2813}, similar to the layered \glspl{tmd}~\cite{1978-Whittingham-PSSC-12-41, 1987-Friend-AP-36-1}.
Although their bandgaps are \SI{\sim 150}{\milli\electronvolt}, doping by intrinsic defects, such as \ch{Ch} vacancies, yields crystals that are far from insulating.
To increase the contrast in conductivity between the bulk and the metallic \gls{tss}, the crystals are often compensated extrinsically.
For example, \ch{Ca} substitution for \ch{Bi} suppresses the self-doped $n$-type conductivity in \ch{Bi2Se3}~\cite{2009-Hor-PRB-79-195208, 2009-Hsieh-N-460-1101}.
Doping can also be used to modulate their magnetic properties, yielding magnetic \glspl{ti} where the \gls{tss} is gapped~\cite{2010-Chen-S-329-659}, for example, by substitution of \ch{Bi} with a paramagnetic transition metal~\cite{2010-Hor-PRB-81-195203, 2019-Tokura-NRP-1-126}.

The intriguing electronic properties of the tetradymite \glspl{ti} have predominantly been investigated using surface sensitive probes in real and reciprocal space (\latin{e.g.}, \gls{sts} and \gls{arpes}), as well as other bulk methods.
In complement to these studies, \gls{nmr} offers the ability to probe their electronic ground state and low-energy excitations through the hyperfine coupling of the nuclear spin probe to the surrounding electrons.
Such a local probe is especially useful when disorder masks sharp reciprocal space features, as is the case in \ch{Bi2Ch3}.
The availability of a useful \gls{nmr} nucleus, however, is usually determined by elemental composition and natural (or enriched) isotopic abundance, as well as the specific nuclear properties such as the gyromagnetic ratio $\gamma$ and, for spin $> 1/2$, the nuclear electric quadrupole moment $Q$.
While the \ch{Bi2Ch3} family naturally contain several \gls{nmr} nuclei~\cite{2013-Nisson-PRB-87-195202, 2014-Koumoulis-AFM-24-1519, 2016-Levin-JPCC-120-25196}, they are either low-abundance or have a large $Q$.
As an alternative, here we use an ion-implanted \gls{nmr} probe at ultratrace concentrations, with detection based on the asymmetric property of radioactive $\beta$-decay, known as \gls{bnmr}~\cite{2015-MacFarlane-SSNMR-68-1}.

A key feature of ion-implanted \gls{bnmr} is the depth resolution
afforded by control of the incident beam energy~\cite{2014-Morris-HI-225-173, 2015-MacFarlane-SSNMR-68-1},
which dictates the stopping distribution of the implanted \gls{nmr} probes.
At \si{\kilo\electronvolt} energies,
the depth can be varied on the nanometer length-scale and,
at the lowest accessible energies,
it may be able to sense the \gls{tss} in the tetradymite \glspl{ti},
which is likely confined to depths \SI{< 1}{\nano\meter} below the surface.
Here one expects Korringa relaxation and Knight shifts from the \gls{tss} electrons,
modified by the phase space restrictions imposed by their chirality.
The magnitude of each will depend on both the \gls{tss} carrier density and
the strength of the coupling to the implanted nuclei.
While the motivation to study the \gls{tss} is strong,
here we report the ``bulk'' response of an implanted \ch{^{8}Li} probe
in two doped \ch{Bi2Ch3} \glspl{ti}.
This is an essential step toward detecting the
\gls{tss}~\cite{2014-MacFarlane-PRB-90-214422, 2019-McFadden-PRB-99-125201},
but it also demonstrates the sensitivity of the implanted \ch{^{8}Li} to
the carriers in such heavily compensated narrow gap semiconductors.

Using \gls{bnmr}, we study two single crystals of doped \ch{Bi2Ch3} --- compensated \gls{bsc} and magnetic \gls{btm} --- each with a beam of highly polarized \ch{^{8}Li^{+}}.
In many respects, \gls{bnmr} is closely related to \gls{musr}, but the radioactive lifetime is much longer, making the frequency range of dynamics it is sensitive to more comparable to conventional \gls{nmr}.
In addition to purely electronic phenomena, in solids containing mobile species, \gls{nmr} is also well known for its sensitivity to low frequency diffusive fluctuations~\cite{1948-Bloembergen-PR-73-679, 1982-Kanert-PR-91-183, 1994-MullerWarmuth-PIR-17-339}, as are often encountered in intercalation compounds.
At ion-implantation energies sufficient to probe the bulk of \gls{bsc} and \gls{btm}, we find evidence for ionic mobility of \ch{^{8}Li^{+}} above \SI{\sim 200}{\kelvin}, likely due to \gls{2d} diffusion in the \gls{vdw} gap.
At low temperature, we find Korringa relaxation and a small temperature dependent negative Knight shift in \gls{bsc}, allowing a detailed comparison with \ch{^{8}Li} in the structurally similar \gls{bts}~\cite{2019-McFadden-PRB-99-125201}.
In \gls{btm}, the effects of the \ch{Mn} moments predominate, but remarkably the signal can be followed through the magnetic transition.
At low temperature, we find a prominent critical peak in the relaxation that is suppressed in a high applied field, and a broad, intense resonance that is strongly shifted.
This detailed characterization of the \ch{^{8}Li} \gls{nmr} response is an important step towards using depth-resolved \gls{bnmr} to study the low-energy properties of the chiral \gls{tss}.

\section{Experiment \label{sec:experiment}}

Doped \gls{ti} single crystals \gls{bsc} and \gls{btm} with nominal stoichiometries \ch{Bi_{1.99}Ca_{0.01}Se_{3}} and \ch{Bi_{1.9}Mn_{0.1}Te_{3}} were grown as described in Refs.~\onlinecite{2009-Hor-PRB-79-195208, 2010-Hor-PRB-81-195203} and magnetically characterized using a Quantum Design \gls{mpms}.
In the \gls{btm}, a ferromagnetic transition was identified at $T_{C} \approx \SI{13}{\kelvin}$, consistent with similar \ch{Mn} concentrations~\cite{2010-Hor-PRB-81-195203, 2013-Watson-NJP-15-103016, 2016-Zimmermann-PRB-94-125205, 2019-Vaknin-PRB-99-220404}.
In contrast, the susceptibility of the \gls{bsc} crystal was too weak to measure accurately,
but the data show no evidence for a Curie tail at low-$T$ that could originate from dilute paramagnetic defects.

\gls{bnmr} experiments were performed at TRIUMF's \gls{isac} facility in Vancouver, Canada.
Detailed accounts of the technique can be found in Refs.~\onlinecite{2015-MacFarlane-SSNMR-68-1, 2019-McFadden-PRB-99-125201}.
A low-energy highly polarized beam of \ch{^{8}Li^{+}} was implanted into the samples mounted in one of two dedicated spectrometers~\cite{2014-Morris-HI-225-173, 2015-MacFarlane-SSNMR-68-1}.
Prior to mounting, the crystals were cleaved in air and affixed to sapphire plates using \ch{Ag} paint (SPI Supplies, West Chester, PA).
The approximate crystal dimensions were \SI{7.8 x 2.5 x 0.5}{\milli\meter} (\gls{bsc}) and \SI{5.3 x 4.8 x 0.5}{\milli\meter} (\gls{btm}).
With the crystals attached, the plates were then clamped to an aluminum holder threaded into an \gls{uhv} helium coldfinger cryostat.
The incident \ch{^{8}Li^{+}} ion beam had a typical flux of \SI{\sim e6}{ions\per\second} over a beam spot \SI{\sim 2}{\milli\metre} in diameter.
At the implantation energies $E$ used here (between \SIrange{1}{25}{\kilo\electronvolt}), \ch{^{8}Li^{+}} stopping profiles were simulated for \num{e5} ions using the \gls{srim} Monte Carlo code (see \Cref{sec:implantation})~\cite{srim}.
For $E > \SI{1}{\kilo\electronvolt}$, a negligible fraction of the \ch{^{8}Li^{+}} stop near enough to the
crystal surface to sense the \gls{tss}.
Most of the data is taken at \SI{20}{\kilo\electronvolt}, where the implantation depth is \SI{\sim 100}{\nano\meter}, and the results thus reflect the bulk behavior.

The probe nucleus \ch{^{8}Li} has nuclear spin $I=2$, gyromagnetic ratio $\gamma / 2 \pi = \SI{6.3016}{\mega\hertz\per\tesla}$, nuclear electric quadrupole moment $Q = \SI[retain-explicit-plus]{+32.6}{\milli\barn}$, and radioactive lifetime $\tau_{\beta} = \SI{1.21}{\second}$.
The nuclear spin is polarized in-flight by collinear optical pumping with circularly polarized light~\cite{2014-Levy-HI-225-165}, yielding a polarization of \SI{\sim 70}{\percent}~\cite{2014-MacFarlane-JPCS-551-012059}.
In each measurement, we alternate the sense of circular polarization
(left and right) of the pumping light,
producing either ``positive'' or ``negative'' helicity in the \ch{^{8}Li} beam
(i.e., its nuclear spin polarization is aligned or counter-aligned with the beam).
Data were collected separately for each helicity, which are usually combined,
but in some cases, are considered independently to reveal ``helicity-resolved'' properties.
While helicity-resolved spectra are useful to elucidate details of resonance lines
(see \Cref{sec:helicities}),
combined spectra are helpful to remove detection systematics
(see \latin{e.g.},~\cite{1983-Ackermann-TCP-31-291, 2015-MacFarlane-SSNMR-68-1}).
The \ch{^{8}Li} polarization was monitored after implantation through the anisotropic radioactive $\beta$-decay, similar to \gls{musr}.
Specifically, the experimental asymmetry $A$ (proportional to the average longitudinal spin-polarization) was measured by combining the rates in two opposed scintillation counters~\cite{1983-Ackermann-TCP-31-291, 2015-MacFarlane-SSNMR-68-1}.
The proportionality constant depends on the experimental geometry and the details of the $\beta$-decay (here, on the order of \num{\sim 0.1}).

\Gls{slr} measurements were performed by monitoring the transient decay of spin-polarization both during and following a pulse of beam lasting several seconds.
During the pulse, the polarization approaches a steady-state value, while after the pulse, it relaxes to essentially zero.
At the edge of the pulse, there is a discontinuity in the slope, characteristic of \gls{bnmr} \gls{slr} spectra (see \latin{e.g.}, \Cref{fig:bsc-slr-spectra}).
Note that unlike conventional \gls{nmr}, no \gls{rf} field is required for the \gls{slr} measurements.
As a result, it is generally more expedient to measure \gls{slr} than the resonance;
however, there is no spectral resolution of the relaxation, which represents the \gls{slr} of all the \ch{^{8}Li}.
The temperature dependence of the \gls{slr} rate was measured at several applied magnetic fields $B_{0}$:
\SI{6.55}{\tesla} parallel to the \ch{Bi2Ch3} trigonal $c$-axis;
and at lower fields \SI{\leq 20}{\milli\tesla} perpendicular to the $c$-axis.
A typical \gls{slr} measurement took \SI{\sim 20}{\minute}.

Resonances were acquired using a \gls{dc} \ch{^{8}Li^{+}} beam and a \gls{cw} transverse \gls{rf} magnetic field $B_{1}$.
In this measurement mode, the \gls{rf} frequency is stepped slowly through the \ch{^{8}Li} Larmor frequency
\begin{equation*} \label{eq:larmor}
   \omega_{0} = 2 \pi \nu_{0} = \gamma B_{0}
\end{equation*}
and the spin of any on-resonance \ch{^{8}Li} is rapidly precessed, resulting in a loss in the average time-integrated $\beta$-decay asymmetry.
The resonance amplitudes are determined by several factors:
the baseline asymmetry (\latin{i.e.}, the time integral of the \gls{slr});
the \gls{rf} amplitude $B_{1}$;
the presence of slow, spectral \ch{^{8}Li} dynamics occurring up to the second timescale (see \latin{e.g.}, Ref.~\onlinecite{2017-McFadden-CM-29-10187});
and, for quadrupole satellite transitions, the relative populations of the magnetic sublevels are somewhat different than conventional pulsed \gls{nmr}~\cite{1990-Slichter-PMR, 2015-MacFarlane-SSNMR-68-1, 2019-McFadden-PRB-99-125201}.
Resonances were recorded over a temperature range of \SIrange{4}{315}{\kelvin} at both high and low magnetic fields.
At high field, the resonance frequency was calibrated against its position in a single crystal \ch{MgO} at \SI{300}{\kelvin}, with the superconducting solenoid persistent.
A single spectrum typically took \SI{\sim 30}{\minute} to acquire.

\section{Results and analysis \label{sec:results}}

\subsection{\ch{Bi2Se3:Ca} \label{sec:results:bsc}}

\begin{figure}
\centering
\includegraphics[width=\columnwidth]{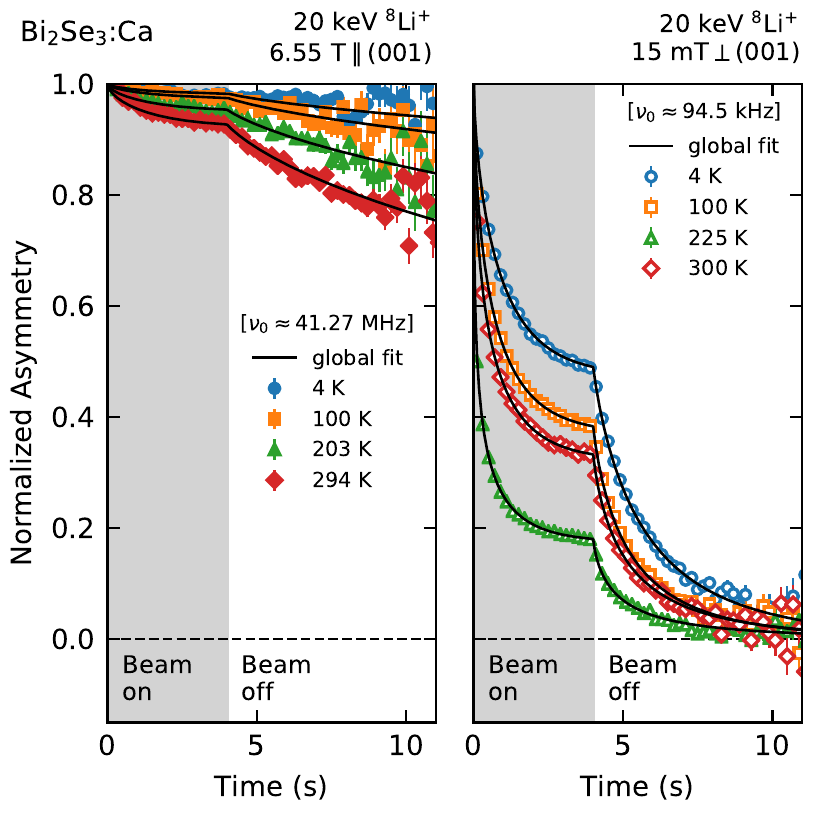}
\caption[
Typical \ch{^{8}Li} \acrlong{slr} spectra in \ch{Ca} doped \ch{Bi2Se3} at high and low magnetic field.
]{ \label{fig:bsc-slr-spectra}
Typical \ch{^{8}Li} \gls{slr} spectra in \ch{Ca} doped \ch{Bi2Se3} at high (left) and low (right) magnetic field with \ch{^{8}Li^{+}} implanted at \SI{20}{\kilo\electronvolt}.
The shaded region indicates the duration of the \ch{^{8}Li^{+}} beam pulse.
The relaxation is strongly field dependent, increasing at lower fields, and it increases non-monotonically with increasing temperature.
The solid black lines show fits to a stretched exponential described in the text.
The initial asymmetry $A_{0}$ from the fits is used to normalize the data which are binned by a factor of \num{20} for clarity.
}
\end{figure}

Typical \ch{^{8}Li} \gls{slr} spectra in \gls{bsc}, at both high and low magnetic field, are shown in \Cref{fig:bsc-slr-spectra}.
To aid comparison, $A(t)$ has been normalized by its initial value $A_{0}$ determined from fits described below.
Clearly, the \gls{slr} is strongly temperature and field dependent.
At low field, the \gls{slr} is very much faster, due to additional relaxation from fluctuations of the host lattice nuclear spins~\cite{2009-Hossain-PRB-79-144518}.
The temperature dependence of the relaxation is non-monotonic, indicating that some of the low frequency fluctuations at $\omega_{0}$ are frozen out at low temperature.

The relaxation is non-exponential at \emph{all} temperatures \emph{and} fields, so the data were fit with the phenomenological stretched exponential.
This approach was also used for \ch{^{8}Li} in \gls{bts}~\cite{2019-McFadden-PRB-99-125201} and in conventional \gls{nmr} of related materials~\cite{2013-Nisson-PRB-87-195202, 2014-Koumoulis-AFM-24-1519, 2016-Levin-JPCC-120-25196}.
Explicitly, for a \ch{^{8}Li^{+}} implanted at time $t^{\prime}$, the spin polarization at time $t > t^{\prime}$ follows:
\begin{equation} \label{eq:strexp}
   R \left ( t, t^{\prime} \right ) = \exp \left \{ - \left [ \lambda \left ( t-t^{\prime} \right ) \right ]^{\beta} \right \},
\end{equation}
where $\lambda \equiv 1/T_{1}$ is the \gls{slr} rate and $0 < \beta \leq 1$ is the stretching exponent.
This is the simplest model that fits the data well with the minimal number of free parameters, for the entire \ch{Bi2Ch3} tetradymite family of \glspl{ti}.

\begin{figure}
\centering
\includegraphics[width=\columnwidth]{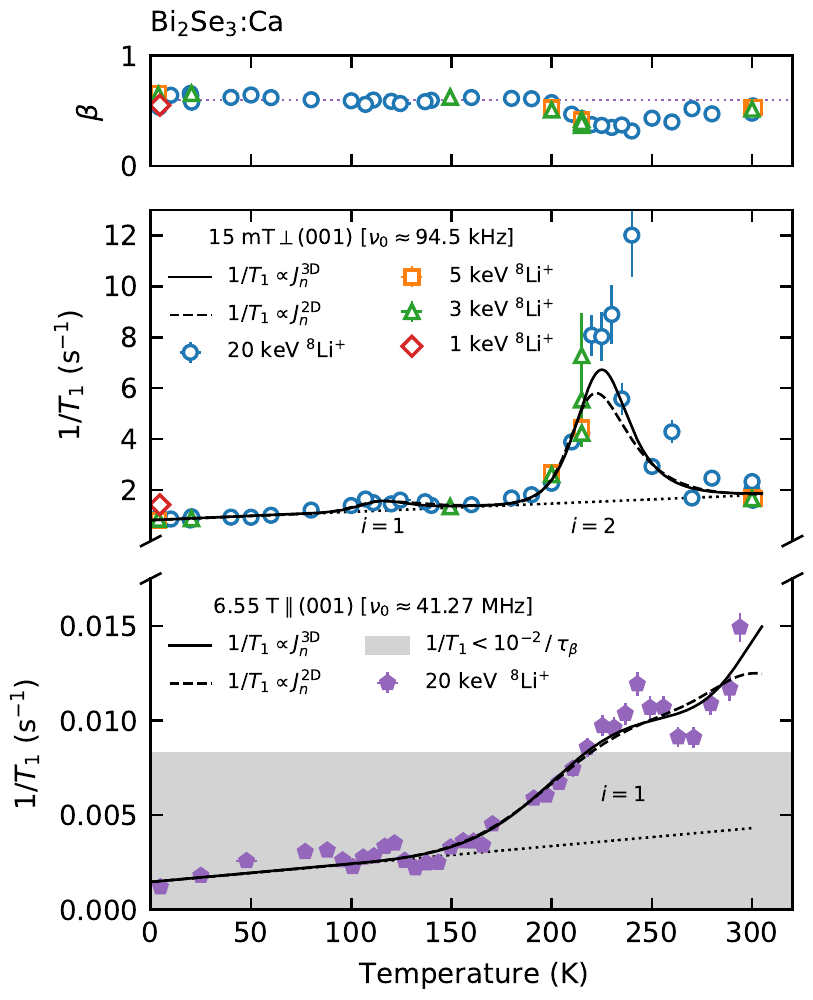}
\caption[
Temperature and field dependence of the \ch{^{8}Li} \acrlong{slr} rate $1/T_{1}$ and stretching exponent $\beta$ in \ch{Ca} doped \ch{Bi2Se3}.
]{ \label{fig:bsc-slr-fits}
Temperature and field dependence of the \ch{^{8}Li} \gls{slr} rate $1/T_{1}$ and stretching exponent $\beta$ in \ch{Ca} doped \ch{Bi2Se3}.
$\beta$ is nearly independent of temperature and field at \num{\sim 0.6} (dotted line), except at low field around the large $1/T_{1}$ peak seen in the bottom panel.
The solid and dashed black lines are global fits to \Cref{eq:rlx}, consisting of a linear $T$-dependence with a non-zero intercept and two \gls{slr} rate peaks, labelled with index $i$.
Independent of the choice of $J_{n}$ used in the analysis, the model captures all the main features of the data.
The $T$-linear contribution to $1/T_{1}$ is shown as the dotted black line.
}
\end{figure}

Using \Cref{eq:strexp} convoluted with the beam pulse, \gls{slr} spectra in \gls{bsc}, grouped by magnetic field $B_{0}$ and implantation energy $E$, were fit simultaneously with a shared common initial asymmetry $A_{0}(B_{0}, E)$.
Note that the statistical uncertainties in the data are strongly time-dependent (see \latin{e.g.}, \Cref{fig:bsc-slr-spectra}), which must be accounted for in the analysis.
Using custom C++ code incorporating the MINUIT minimization routines~\cite{1975-James-CPC-10-343} implemented within the ROOT data analysis framework~\cite{1997-Brun-NIMA-389-81}, we find the global least-squares fit for each dataset.
The fit quality is good ($\tilde{\chi}_{\mathrm{global}}^{2} \approx 1.02$) and a subset of the results are shown in \Cref{fig:bsc-slr-spectra} as solid black lines.
The large values of $A_{0}$ extracted from the fits (\SI{\sim 10}{\percent} for $B_{0} = \SI{6.55}{\tesla}$ and \SI{\sim 15}{\percent} for $B_{0} = \SI{15}{\milli\tesla}$) are consistent with the full beam polarization, with no missing fraction.
The fit parameters are plotted in \Cref{fig:bsc-slr-fits}, showing agreement with the qualitative observations above.

We now consider a model for the temperature and field dependence of $1/T_{1}$.
We interpret the local maxima in $1/T_{1}$ in \Cref{fig:bsc-slr-fits} as \gls{bpp} peaks~\cite{1948-Bloembergen-PR-73-679}, caused by a fluctuating field coupled to the \ch{^{8}Li} nuclear spin with a characteristic rate that sweeps through $\omega_{0}$ at the peak temperature~\cite{1948-Bloembergen-PR-73-679, 1979-Richards-TCP-15-141, 1988-Beckmann-PR-171-85}.
Potential sources of the fluctuations are discussed below.
The rate peaks are superposed on a smooth background that is approximately linear, reminiscent of Korringa relaxation in metals~\cite{1950-Korringa-P-16-601, 1990-Slichter-PMR}.
This is surprising, since \gls{bsc} is a semiconductor, but it is similar to \gls{bts}~\cite{2019-McFadden-PRB-99-125201}.
We discuss this point further in \Cref{sec:discussion:electronic}.

From this, we adopt the following model for the total \gls{slr} rate:
\begin{equation} \label{eq:rlx}
   1/T_{1} = a + b T + \sum_{i} c_{i} \left ( J_{1,i} + 4J_{2,i} \right ) .
\end{equation}
In \Cref{eq:rlx}, the first two terms account for the $T$-linear contribution with a finite intercept $a$, while the remaining terms describe the $i^{\mathrm{th}}$ $1/T_{1}$ peak in terms of a coupling constant $c_{i}$ (proportional to the mean-squared transverse fluctuating field) and the $n$-quantum \gls{nmr} spectral density functions $J_{n,i}$~\cite{1988-Beckmann-PR-171-85}.
In general, $J_{n,i}$ is frequency dependent and peaked at a temperature where the fluctuation rate matches $\sim n \omega_{0}$.
While the precise form of $J_{n,i}$ is not known \latin{a priori}, the simplest expression, obtained for isotropic \gls{3d} fluctuations, has a Debye (Lorentzian) form~\cite{1948-Bloembergen-PR-73-679, 1988-Beckmann-PR-171-85}:
\begin{equation} \label{eq:j3d}
   J_{n}^{\mathrm{3D}} = \frac{\tau_{c}}{1 + \left (n \omega_{0} \tau_{c} \right )^{2}} ,
\end{equation}
where $\tau_{c}$ is the (exponential) correlation time of the fluctuations.
Alternatively, when the fluctuations are \gls{2d} in character, as might be anticipated for such a layered crystal,
$J_{n}$ may be described by the empirical expression~\cite{1979-Richards-TCP-15-141, 1994-Kuchler-SSI-70-434}:
\begin{equation} \label{eq:j2d}
   J_{n}^{\mathrm{2D}} = \tau_{c} \ln \left ( 1 + \left ( n \omega_{0} \tau_{c} \right )^{-2} \right ).
\end{equation}
For both \Cref{eq:j3d,eq:j2d}, we assume that $\tau_{c}$ is thermally activated, following an Arrhenius dependence:
\begin{equation} \label{eq:arrhenius}
   \tau_{c}^{-1} = \tau_{0}^{-1} \exp \left ( - \frac{ E_{A} }{ k_{B} T } \right ),
\end{equation}
where $E_{A}$ is the activation energy, $\tau_{0}$ is a prefactor, $k_{B}$ is the Boltzmann constant.
If the fluctuations are due to \ch{^{8}Li^{+}} hopping, $\tau_{c}^{-1}$ is the site-to-site hop rate.

Using the above expressions, we fit the $1/T_{1}$ data using a global procedure wherein the kinetic parameters (\latin{i.e.}, $E_{A, i}$ and $\tau_{0, i}^{-1}$) are shared at all the different $\omega_{0}$.
This was necessary to fit the data at \SI{6.55}{\tesla} where the relaxation is very slow.
For comparison, we applied this procedure using both $J_{n}^{\mathrm{3D}}$ and $J_{n}^{\mathrm{2D}}$ and the fit results are shown in \Cref{fig:bsc-slr-fits} as solid ($J_{n}^{\mathrm{3D}}$) and dashed ($J_{n}^{\mathrm{2D}}$) lines, clearly capturing the main features of the data.
The analysis distinguishes two processes, $i = 1, 2$ in \Cref{eq:rlx}:
one $(i = 1)$ that onsets at lower temperature with a shallow Arrhenius slope of \SI{\sim 0.1}{\electronvolt}
that yields the weaker peaks in $1/T_1$ at both fields;
and a higher barrier process $(i = 2)$ with an $E_{A}$ of \SI{\sim 0.4}{\electronvolt} that
yields the more prominent peak in the low field relaxation,
while the corresponding high field peak must lie above the accessible temperature range.
The resulting fit parameters are given in \Cref{tab:bsc-slr-fits}.
We discuss the results in \Cref{sec:discussion:dynamics}.

\begin{table}
\centering
\caption[
Arrhenius parameters obtained from the analysis of the temperature dependence of $1/T_{1}$ in \ch{Ca} doped \ch{Bi2Se3}.
]{ \label{tab:bsc-slr-fits}
Arrhenius parameters in \Cref{eq:arrhenius} obtained from the analysis of the temperature dependence of $1/T_{1}$ in \ch{Ca} doped \ch{Bi2Se3} shown in \Cref{fig:bsc-slr-fits}.
The two processes giving rise to the rate peaks are labelled with index $i$.
Good agreement is found between the $E_{A}$s determined using the spectral density functions $J_{n}$ for \gls{2d} and \gls{3d} fluctuations [\Cref{eq:j2d,eq:j3d}].
}
\begin{tabular}{c S S S S}
\toprule
& \multicolumn{2}{c}{$i = 1$} & \multicolumn{2}{c}{$i = 2$} \\
$J_{n}$ & {$\tau_{0}^{-1}$ (\SI{e10}{\per\second})} & {$E_{A}$ (\si{\electronvolt})} & {$\tau_{0}^{-1}$ (\SI{e14}{\per\second})} & {$E_{A}$ (\si{\electronvolt})} \\
\midrule
3D & 8.4 \pm 2.7 & 0.113 \pm 0.005 & 7 \pm 5 & 0.395 \pm 0.015 \\
2D & 9 \pm 3 & 0.106 \pm 0.005 & 110 \pm 90 & 0.430 \pm 0.016 \\
\bottomrule
\end{tabular}
\end{table}

\begin{figure}
\centering
\includegraphics[width=\columnwidth]{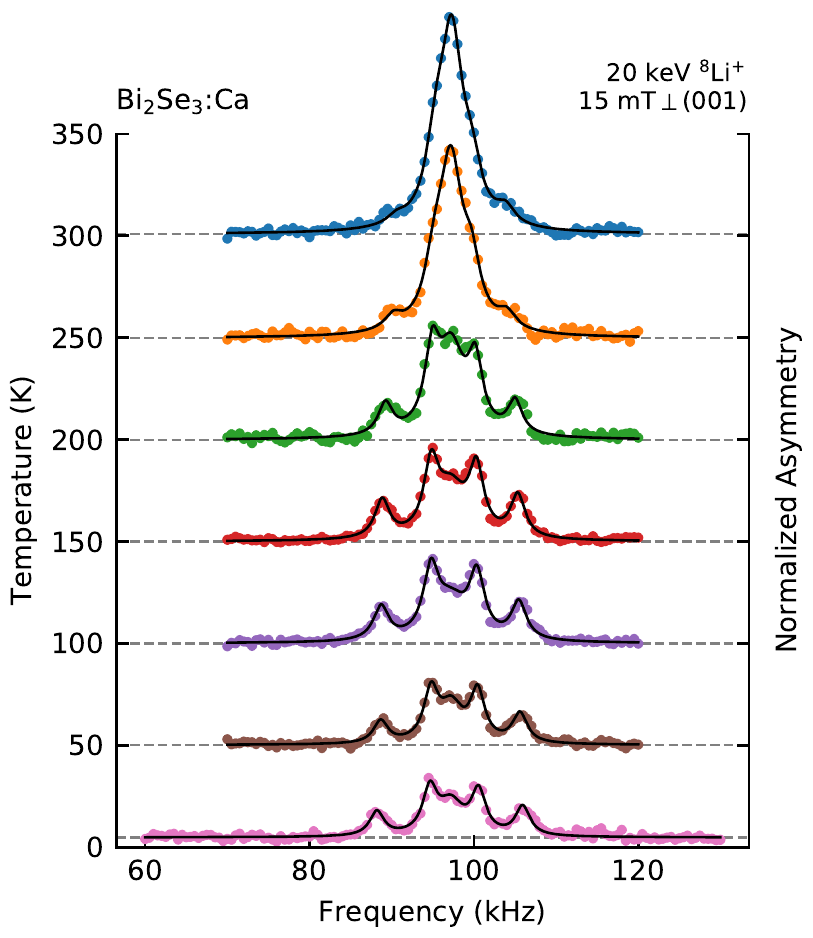}
\caption[
\ch{^{8}Li} resonance spectra in \acrlong{bsc} at low magnetic field.
]{ \label{fig:bsc-1f-spectra-lf}
\ch{^{8}Li} resonance spectra in \gls{bsc} at low magnetic field.
The vertical scale is the same for all spectra;
they have been normalized to account for changes in intensity due to \gls{slr}~\cite{2009-Hossain-PB-404-914}, with their baselines (shown as dashed grey lines) shifted to match the temperature.
The spectra consist of a small and nearly temperature-independent quadrupole split pattern, centred about an unsplit Lorentzian line, whose amplitude grows above \SI{\sim 150}{\kelvin}.
Note the quadrupole pattern of an integer spin nucleus like \ch{^{8}Li}, has no central satellite (main line).
The solid black lines are fits to a sum of this Lorentzian plus and $2I = 4$ quadrupole satellites (see text).
}
\end{figure}

We now turn to the \ch{^{8}Li} resonances, with typical spectra shown in \Cref{fig:bsc-1f-spectra-lf}.
As anticipated for a non-cubic crystal, the spectrum is quadrupole split, confirmed unambiguously by the helicity-resolved spectra (see \Cref{fig:bsc-1f-spectra-helicities} in \Cref{sec:helicities}).
This splitting, on the order of a few \si{\kilo\hertz}~\footnote{Note that the large $A_{0}$ down to low field precludes \ch{^{8}Li^{+}} sites with very large \glspl{efg}.},
is determined by the \gls{efg} and is a signature of the crystallographic \ch{^{8}Li} site.
Besides this, an unsplit component is also apparent, very close to (within \SI{\sim 100}{\hertz}) the centre-of-mass of the four satellites.
At low temperature, the ``central'' and split components are nearly equal, but as the temperature is raised, the unsplit line grows to dominate the spectrum, accompanied by a slight narrowing.

The scale of the quadrupole splitting is determined by the product of the principal component of the \gls{efg} tensor $eq$ with the nuclear electric quadrupole moment $eQ$.
We quantify this with a conventional definition of the quadrupole frequency (for $I=2$)~\cite{1957-Cohen-SSP-5-321}:
\begin{equation*}
   \nu_{q} = \frac{e^{2} q Q}{8 h}.
\end{equation*}
In high field, a first order perturbation treatment of the quadrupole interaction is sufficient to obtain accurate satellite positions.
However, at low field, where $\nu_{q} / \nu_{0} \approx \SI{6}{\percent}$, second order terms are required~\cite{1957-Cohen-SSP-5-321, 1990-Taulelle-NASC-393}.
Based on the change in satellite splittings by a factor \num{2} in going from $B_{0} \parallel c$ to $B_{0} \perp c$, we assume the asymmetry parameter of the \gls{efg} $\eta = \num{0}$ (\latin{i.e.}, the \gls{efg} is axially symmetric).
This is reasonable based on likely interstitial sites for \ch{^{8}Li^{+}}~\cite{2019-McFadden-PRB-99-125201}.
Pairs of helicity-resolved spectra were fit with $\nu_{0}$ and $\nu_{q}$ as shared free parameters, in addition to linewidths and amplitudes.
As the difference between the frequency of the unsplit line and the center of the quadrupole split pattern was too small to measure accurately,
the fits were additionally constrained to have the same central frequency $\nu_{0}$.
This is identical to the approach used for \gls{bts}~\cite{2019-McFadden-PRB-99-125201}.
A subset of the results (after recombining the two helicities) are shown in \Cref{fig:bsc-1f-spectra-lf} as solid black lines.

\begin{figure}
\centering
\includegraphics[width=\columnwidth]{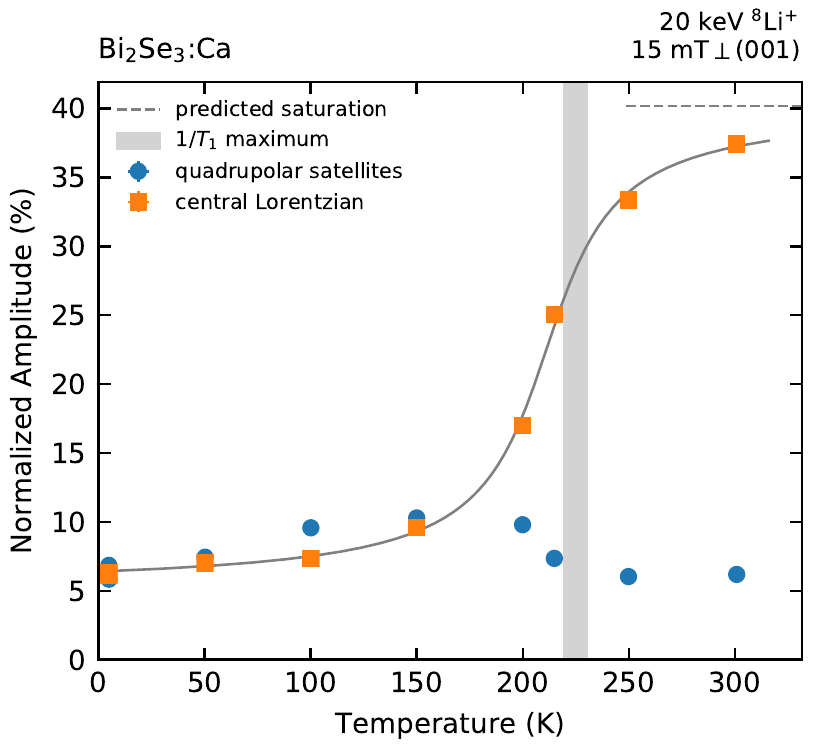}
\caption[
The resonance amplitude as a function of temperature in \ch{Ca} doped \ch{Bi2Se3} at \SI{15}{\milli\tesla}.
]{ \label{fig:bsc-1f-fits-amp}
The resonance amplitude as a function of temperature in \ch{Ca} doped \ch{Bi2Se3} at \SI{15}{\milli\tesla}.
While the amplitude of the satellite lines are nearly temperature independent, the central component increases substantially above \SI{150}{\kelvin}, plateauing on the high-$T$ side of the $1/T_{1}$ maximum (grey band).
The solid line is a guide, while the dashed line indicates the estimated saturation value for the Lorentizan component.
}
\end{figure}

The main result is the strong temperature dependence of the resonance amplitude shown in \Cref{fig:bsc-1f-fits-amp}.
While the satellite amplitudes are nearly temperature independent, the central component increases substantially above \SI{150}{\kelvin}, tending to plateau above the $1/T_{1}$ peak.
The other parameters are quite insensitive to temperature.
Typical linewidths (\latin{i.e.}, \acrlong{fwhm}) are \SI{\sim 2.2}{\kilo\hertz} for the satellites and \SI{\sim 3.8}{\kilo\hertz} for the central component.
The quadrupole frequency $\nu_{q} \approx \SI{5.5}{\kilo\hertz}$ varies weakly, increasing slightly as temperature is lowered.

We also measured resonances at room and base temperature in high field (\SI{6.55}{\tesla}) where \ch{^{8}Li} is sensitive to small magnetic shifts.
From the fits, we use $\nu_{0}$ to calculate the raw relative frequency shift $\delta$ in parts per million (\si{\ppm}) using:
\begin{equation} \label{eq:shift}
   \delta = 10^{6} \left ( \frac{\nu_{0} - \nu_{\ch{MgO}} }{ \nu_{\ch{MgO}} } \right ) ,
\end{equation}
where $\nu_{\ch{MgO}}$ is the reference frequency position in \ch{MgO} at \SI{300}{\kelvin}.
The shifts are small:
\SI[retain-explicit-plus]{+12 \pm 2}{\ppm} at room temperature and \SI{-17 \pm 3}{\ppm} at \SI{5}{\kelvin}, the latter considerably smaller in magnitude than in \gls{bts}~\cite{2019-McFadden-PRB-99-125201}.
Because \ch{^{8}Li} \gls{nmr} shifts are generally so small, it is essential to account for the demagnetization field of the sample itself.
From $\delta$, the corrected shift $K$ is obtained by the \gls{cgs} expression~\cite{2008-Xu-JMR-191-47}:
\begin{equation}
   K = \delta + 4 \pi \left ( N - \frac{1}{3} \right ) \chi_{v}
\end{equation}
where $N$ is the dimensionless demagnetization factor that depends only on the shape of the sample and $\chi_{v}$ is the dimensionless (volume) susceptibility.
For a thin film, $N$ is \num{1}~\cite{2008-Xu-JMR-191-47}, but for the thin platelet crystals used here, we estimate $N$ is on the order of \num{\sim 0.8}, treating them as oblate ellipsoids~\cite{1945-Osborn-PR-67-351}.
For the susceptibility, we take the average of literature values reported for pure \ch{Bi2Se3}~\cite{1958-Matyas-CJP-8-309, 2003-Kulbachinskii-PB-329-1251, 2015-Pafinlov-JPCM-27-456002, 2016-Chong-JAC-686-245}, giving $\chi_{v}^{\mathrm{CGS}} \approx \SI{-2.4e-6}{\emu\per\centi\meter\cubed}$.
Note that we have excluded several reports~\cite{2008-Janicek-PB-403-3553, 2012-Young-PRB-86-075137, 2014-Zhao-NM-13-580} whose results disagree by an order of magnitude from those predicted by Pascal's constants~\cite{2008-Bain-JCE-85-532}.
Applying the correction for \gls{bsc} yields $K$s of \SI[retain-explicit-plus]{-2 \pm 2}{\ppm} and \SI{-31 \pm 3}{\ppm} at room and base temperature, respectively.
We discuss this below in \Cref{sec:discussion:electronic}.

\subsection{\ch{Bi2Te3:Mn} \label{sec:results:btm}}

\begin{figure}
\centering
\includegraphics[width=\columnwidth]{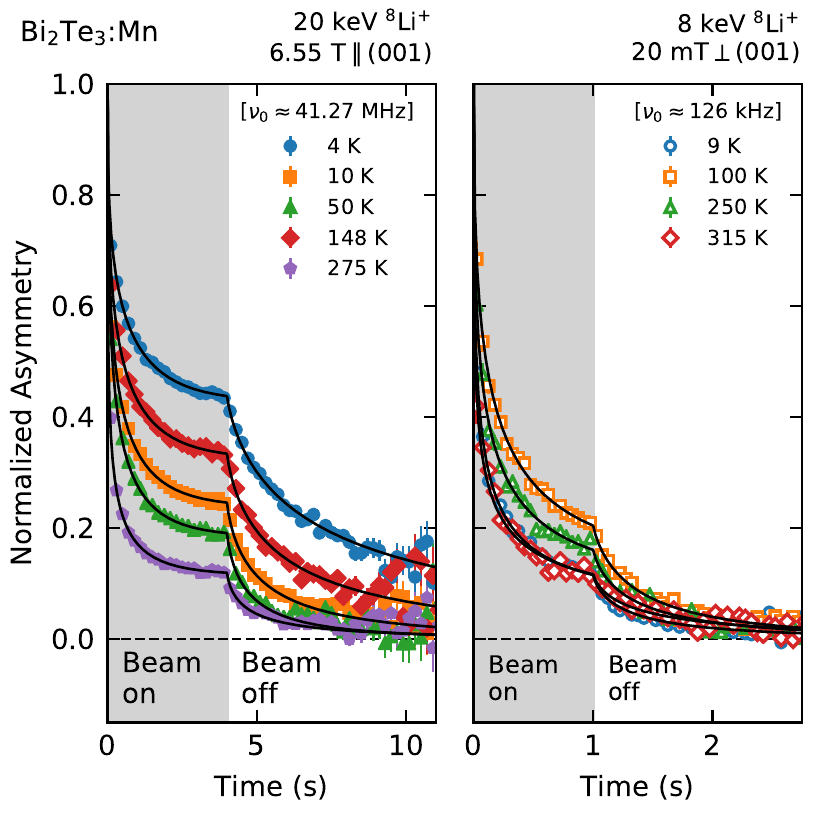}
\caption[
Typical \ch{^{8}Li} \acrlong{slr} spectra in \ch{Mn} doped \ch{Bi2Te3} at high and low magnetic field.
]{ \label{fig:btm-slr-spectra}
Typical \ch{^{8}Li} \gls{slr} spectra in \ch{Mn} doped \ch{Bi2Te3} at high (left) and low (right) magnetic field for \ch{^{8}Li^{+}} implantation energies of \SI{20}{\kilo\electronvolt} and \SI{8}{\kilo\electronvolt}, respectively.
The shaded region denotes the duration of the \ch{^{8}Li^{+}} beam pulse.
The \gls{slr} is substantial and orders of magnitude faster than in \gls{bsc} (see \Cref{fig:bsc-slr-spectra}) at high field.
The field dependence to the \gls{slr} is much weaker than in the nonmagnetic tetradymites.
The solid black lines are fits to a stretched exponential convoluted with the \ch{^{8}Li^{+}} beam pulse as described in the text.
The initial asymmetry $A_{0}$ from the fit is used to normalize the spectra.
The high and low field spectra have been binned for by factors of \num{20} and \num{5}, respectively.
}
\end{figure}

Typical \ch{^{8}Li} \gls{slr} spectra at high and low field in the magnetically doped \gls{btm} are shown in \Cref{fig:btm-slr-spectra}.
In contrast to nonmagnetic \gls{bsc}, the relaxation at high field is fast, typical of paramagnets with unpaired electron spins~\cite{2015-MacFarlane-PRB-92-064409, 2016-Cortie-PRL-116-106103,2011-Song-PRB-84-054414}.
The fast high field rate produces a much less pronounced field dependence.
At low field, the \gls{slr} rate is peaked at low temperature.
The relaxation is also non-exponential and fits well using \Cref{eq:strexp}, with a stretching exponent systematically smaller than in the nonmagnetic \gls{bsc} or \gls{bts}~\cite{2019-McFadden-PRB-99-125201}.
We analyzed the data with the same global approach, obtaining good quality fits ($\tilde{\chi}_{\mathrm{global}}^{2} \approx 1.01$) demonstrated by the solid black lines in \Cref{fig:btm-slr-spectra}.
The shared values of $A_{0}$ from the fits are large (\SI{\sim 10}{\percent} for $B_{0} = \SI{6.55}{\tesla}$ and \SI{\sim 15}{\percent} for $B_{0} = \SI{20}{\milli\tesla}$), consistent with the full beam polarization, implying that there is remarkably no magnetic wipeout from very fast relaxation~\cite{2015-MacFarlane-PRB-92-064409}, even at low field.

\begin{figure}
\centering
\includegraphics[width=\columnwidth]{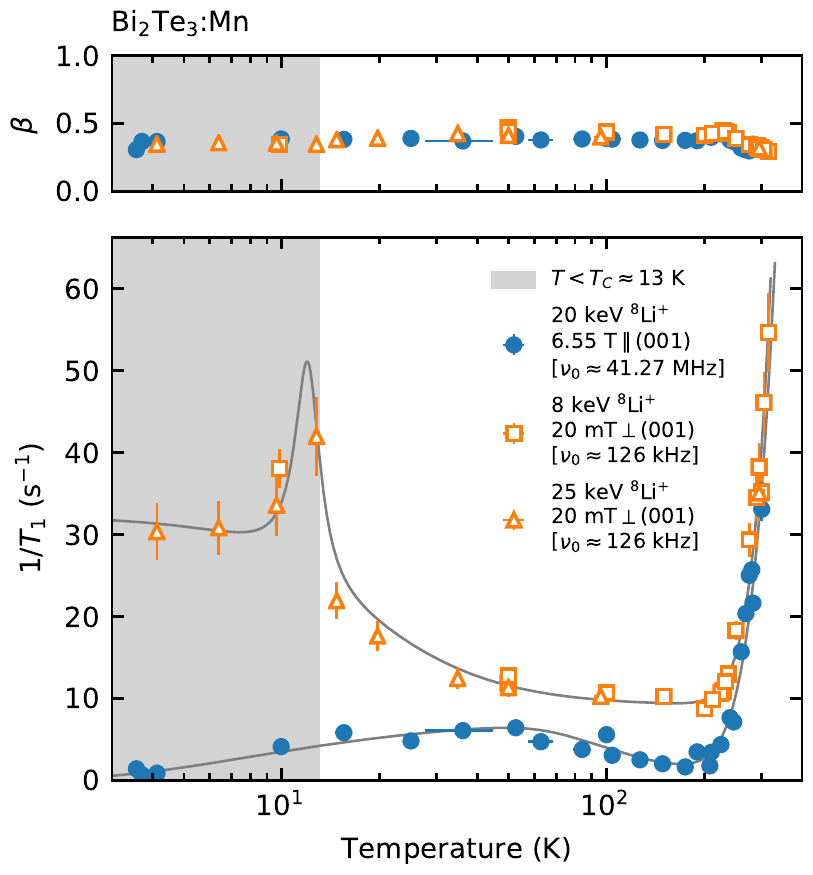}
\caption[
Temperature dependence of the \ch{^{8}Li} \acrlong{slr} rate $1/T_{1}$ in \ch{Mn} doped \ch{Bi2Te3} at high and low field.
]{ \label{fig:btm-slr-fits}
Temperature dependence of the \ch{^{8}Li} \gls{slr} rate $1/T_{1}$ in \ch{Mn} doped \ch{Bi2Te3} at high and low field.
At low field, $1/T_{1}$ shows a critical peak at the ferromagnetic transition at $T_{C} \approx \SI{13}{\kelvin}$, as the \ch{Mn} spin fluctuations freeze out.
Above \SI{200}{\kelvin}, the \gls{slr} rate increases exponentially in manner nearly independent of applied field.
The solid grey lines are drawn to guide the eye.
}
\end{figure}

The fit parameters are shown in \Cref{fig:btm-slr-fits}.
At all temperatures, especially at high field, the \gls{slr} rate $1/T_{1}$ is orders of magnitude larger than in the nonmagnetic analogs.
No clear $1/T_{1}$ \gls{bpp} peaks can be identified between \SIrange[range-units=single,range-phrase=--]{100}{300}{\kelvin};
however, in the low field data, a critical divergence is evident at the magnetometric transition at about \SI{13}{\kelvin}.
In high field, this feature is largely washed out, with a remnant peak near \SI{50}{\kelvin}.
Above \SI{200}{\kelvin}, the \gls{slr} rate increases very rapidly and is well-described by $1/T_{1} \propto \exp [ - E_{A} / ( k_{B} T )]$,
with $E_{A} \approx \SI{0.2}{\electronvolt}$ at both fields.
We discuss this below in \Cref{sec:discussion:dynamics}.

\begin{figure}
\centering
\includegraphics[width=\columnwidth]{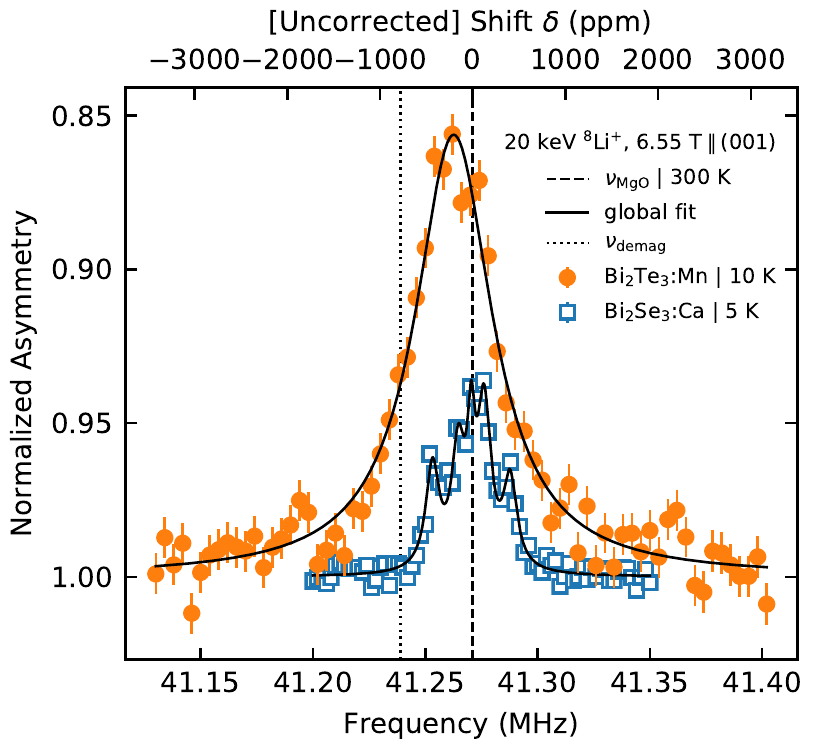}
\caption[
Typical \ch{^{8}Li} resonances in \ch{Mn} doped \ch{Bi2Te3} and \ch{Ca} doped \ch{Bi2Se3} at high magnetic field.
]{ \label{fig:btm-1f-spectra}
Typical \ch{^{8}Li} resonances in \ch{Mn} doped \ch{Bi2Te3} and \ch{Ca} doped \ch{Bi2Se3} at high magnetic field.
The vertical scale is the same for both spectra;
they have been normalized to account for changes in intensity and baseline~\cite{2009-Hossain-PB-404-914}.
The lineshape in the magnetic \gls{btm} is well-described by a broad Lorentzian (solid black line) with no quadrupolar splitting.
A large negative shift is also apparent for \gls{btm} with respect to the reference frequency in \ch{MgO} (vertical dashed line).
The dotted vertical line indicates the expected resonance position due to demagnetization, revealing a large positive hyperfine field (\SI{\sim 38}{\gauss}) at \SI{5}{\kelvin} in the magnetic state.
}
\end{figure}

In contrast to \gls{bsc} and \gls{bts}~\cite{2019-McFadden-PRB-99-125201}, the resonance in \gls{btm} consists of a single broad Lorentzian with none of the resolved fine structure (see \Cref{fig:btm-1f-spectra}).
Surprisingly, the very broad line has significant intensity, dwarfing the quadrupole pattern in \gls{bsc} in both width and amplitude.
In addition, there is a large negative shift at \SI{10}{\kelvin}.
At room temperature, the line is somewhat narrower, and the shift is reduced in magnitude.
Quantitative results from Lorentzian fits are summarized in \Cref{tab:btm-1f-fits}.

\begin{table}
\centering
\caption[
Results from the analysis of the \ch{^{8}Li} resonance in \acrlong{btm} at high and low temperature with $B_{0} = \SI{6.55}{\tesla} \parallel (001)$.
]{ \label{tab:btm-1f-fits}
Results from the analysis of the \ch{^{8}Li} resonance in \gls{btm} at high and low temperature with $B_{0} = \SI{6.55}{\tesla} \parallel (001)$.
The (bulk) magnetization $M$ measured with $\SI{1.0}{\tesla} \parallel (001)$ is included for comparison.
}
\begin{tabular}{S S S S[retain-explicit-plus] S}
\toprule
{$T$ (\si{\kelvin})} & {$\tilde{A}$ (\si{\percent})} & {\acrshort{fwhm} (\si{\kilo\hertz})} & {$\delta$ (\si{\ppm})} & {$M$ (\si{\emu\per\centi\meter\cubed})} \\
\midrule
294 & 33 \pm 5     & 16.2 \pm 1.2 &  +10 \pm 9 &  0.076 \\
 10 & 14.4 \pm 0.8 & 41.7 \pm 1.6 & -206 \pm 12 & 7.698 \\
\bottomrule
\end{tabular}
\end{table}

\section{Discussion \label{sec:discussion}}

The \ch{^{8}Li} \gls{nmr} properties of nonmagnetic \gls{bsc} are quite similar to previous measurements in \gls{bts}~\cite{2019-McFadden-PRB-99-125201}.
The resonance spectra show a similar splitting ($\nu_q$ is about \SI{25}{\percent} smaller in \gls{bsc}), indicating a similar site for \ch{^{8}Li}.
The resemblance of the spectra extends to the detailed temperature dependence, including the growth of the unsplit line approaching room temperature.
Surprisingly, the \gls{bsc} spectra are better resolved than \gls{bts}, implying a higher degree of order, despite the \ch{Ca} doping.
This is also evident in the \gls{slr}, with a stretching exponent $\beta$ closer to unity in \gls{bsc} than in \gls{bts}.
This likely reflects additional disorder in \gls{bts} from \ch{Bi/Te} anti-site defects~\cite{2012-Scanlon-AM-24-2154} that are much more prevalent than for \ch{Bi/Se}, due to the difference in radii and electronegativity.
The sharp quadrupolar pattern indicates a well-defined crystallographic \ch{^{8}Li^{+}} site, and the corresponding small \gls{efg} suggests it is in the \gls{vdw} gap.
\Gls{dft} calculations of the \gls{efg} may enable a precise site assignment.
The high field \gls{slr} is also similar to \gls{bts}:
it is slow and near the lower limit measurable due to the finite \ch{^{8}Li} lifetime $\tau_\beta$ and comparable to the \gls{vdw} metal \ch{NbSe2}~\cite{2006-Wang-PB-374-239}, where the carrier concentration is much higher, but significantly slower than the \gls{ti} alloy \ch{Bi_{1-x}Sb_{x}}~\cite{2014-MacFarlane-PRB-90-214422}.
The low field enhancement of $1/T_1$ is also similar, so it cannot be essentially related to the dilute \ch{^{125}Te} moments that are absent in \gls{bsc}, but probably determined primarily by the \SI{100}{\percent} abundant \ch{^{209}Bi}.
From such a detailed similarity, it is clear that a quantitative comparison with \gls{bts} and other \gls{vdw} materials will be useful.

With these similarities in mind, the rest of the discussion is organized as follows: in \Cref{sec:discussion:dynamics},
we consider evidence of mobility of the \ch{^{8}Li^{+}} ion;
in \Cref{sec:discussion:electronic}, electronic effects at low temperature in \gls{bsc};
and the magnetic properties of \gls{btm} in \Cref{sec:discussion:magnetic}.

\subsection{Dynamics of the \ch{Li^{+}} ion \label{sec:discussion:dynamics}}

In \gls{bts}, we considered if the evolution of the spectrum with temperature (similar to \Cref{fig:bsc-1f-spectra-lf}) was the result of a site change transition from a meta-stable quadrupolar \ch{^{8}Li^{+}} site to a lower energy site with very small \gls{efg} at higher temperature~\cite{2019-McFadden-PRB-99-125201}, similar to elemental \ch{Nb}~\cite{2009-Parolin-PRB-80-174109}.
We now consider an alternative explanation; namely, dynamic averaging of the quadrupolar interaction due to \ch{^{8}Li^{+}} motion.
Examples of this are found in conventional \ch{^{7}Li} \gls{nmr}, where, unlike \ch{^{8}Li}, the $I = 3/2$ quadrupole spectrum has a main line (the $m = \pm 1/2$ satellite) overlapping the averaged resonance~\cite{2012-Galven-CM-24-3335, 2012-Indris-JPCC-116-14243}.
Dynamic averaging is suggested by the onset near, but below, the \gls{slr} rate peak (see \Cref{fig:bsc-1f-fits-amp}).
However, for hopping between equivalent interstitial sites (probably the quasi-octahedral Wyckoff $3b$ site in the \gls{vdw} gap --- see Figure~12 in Ref.~\onlinecite{2019-McFadden-PRB-99-125201}),
one does not expect that the \gls{efg} will average to a value near zero (required to explain the unsplit line).
A point charge estimate reveals the quasi-tetrahedral (Wyckoff $6c$) site, thought to be the saddle point in the potential for \ch{Li^{+}} between adjacent $3b$ sites~\cite{2016-Gosalvez-PRB-93-075429}, has an \gls{efg} of opposite sign to the $3b$ site.
If $6c$ is instead a shallow minimum, the \ch{^{8}Li^{+}} residence time there may be long enough that the \gls{efg} averages to near zero.
In the fast motion limit at higher temperatures, one would then expect the quadrupole splitting to re-emerge when the residence time in the $6c$ ``transition'' site becomes much shorter~\cite{2012-Indris-JPCC-116-14243}.

We now examine the two kinetic processes causing the $1/T_{1}$ peaks in \gls{bsc}.
It is surprising to find two distinct thermally activated processes sweeping through the \gls{nmr} frequency, especially since only a single process was found in \gls{bts}~\cite{2019-McFadden-PRB-99-125201}.
First, we consider the weaker feature, the low temperature $(i=1)$ peaks.
In layered materials, small intercalates can undergo highly localized motion at relatively low temperatures below the onset of free diffusion~\cite{1982-Kanert-PR-91-183, 1994-MullerWarmuth-PIR-17-339}.
Such local motion may be the source of the \gls{slr} rate peak, but it is quite ineffective at narrowing the resonance, consistent with the absence of any lineshape changes in the vicinity of the $i=1$ peaks.
Caged local motion is usually characterized by a small activation barrier, comparable to the \SI{\sim 0.1}{\electronvolt} observed here.
Similar phenomena have been observed at low temperature, for example, in neutron activated \ch{^{8}Li} \gls{bnmr} of \ch{Li} intercalated graphite, \ch{LiC12}~\cite{1989-Schirmer-SM-34-589, 1995-Schirmer-ZNA-50-643}.
It is not clear why such motion would be absent in \gls{bts}~\cite{2019-McFadden-PRB-99-125201}, which has a larger \gls{vdw} gap than \gls{bsc} (\SI{2.698}{\angstrom} vs.\ \SI{2.568}{\angstrom}).
Alternatively, this feature in the relaxation may have an electronic origin,
perhaps related to the emergent low-$T$ magnetism in \ch{MoTe2} observed
by \gls{musr}~\cite{2018-Guguchia-SA-4-eaat3672} and \ch{^{8}Li} \gls{bnmr}~\cite{Krieger-tbp}.

In contrast, the \gls{slr} rate peak above \SI{200}{\kelvin} $(i = 2)$ is almost certainly due to \gls{efg} fluctuations caused by stochastic \ch{^{8}Li^{+}} motion.
From the data, we cannot conclude that this is long-range diffusion, but the room temperature \ch{Li^{+}} intercalability of \ch{Bi2Se3}~\cite{1988-Paraskevopoulos-MSEB-1-147, 1989-Julien-SSI-36-113, 2010-Bludska-JSSC-183-2813} suggests it is.
Its barrier, on the order of \SI{\sim 0.4}{\electronvolt}, is comparable to other \gls{vdw} gap layered ion conductors, but it is about twice as high as in \gls{bts}~\cite{2019-McFadden-PRB-99-125201}, possibly a result of the \ch{Se} (rather than \ch{Te}) bounded \gls{vdw} gap, which provides less space between neighbouring \glspl{ql}.

We now consider the Arrhenius law prefactors $\tau_{0}^{-1}$, that, for ionic diffusion, are typically in the range \SIrange[range-phrase=--,range-units=single]{e12}{e14}{\per\second}.
For the low-$T$ process ($i = 1$), independent of the form of $J_{n}$ (see \Cref{tab:bsc-slr-fits}), $\tau_{0}^{-1} \approx \SI{9e10}{\per\second}$ is unusually low.
In contrast, for the high-$T$ ($i = 2$) process, it is much larger and depends strongly on $J_{n}$.
For \gls{3d} diffusion, it is in the expected range, while the \gls{2d} model yields an extremely large value, \SI{\sim e16}{\per\second},
in the realm of prefactor anomalies~\cite{1983-Villa-SSI-9-1421} and opposite to the small value expected for
low dimensional diffusion~\cite{1978-Richards-SSC-25-1019}.
Similar behaviour was observed recently in \ch{^{7}Li} \gls{nmr} of \ch{LiC6}~\cite{2013-Langer-PRB-88-094304},
where surprisingly, $J_{n}^{\mathrm{2D}}$ was concluded to be less appropriate than $J_{n}^{\mathrm{3D}}$,
suggesting that \ch{Li} motion in the \gls{vdw} gap is not as ideally \gls{2d} as might be expected.
In \gls{bsc}, the anomaly may be related to local dynamics that onset at lower $T$, imparting some \gls{3d} character to the motion.

Given the evidence for long-range \ch{Li^{+}} motion in \gls{bsc} and \gls{bts}~\cite{2019-McFadden-PRB-99-125201}, the absence of a relaxation peak in \gls{btm} may seem unexpected.
Both \ch{Ca^{2+}} and \ch{Mn^{2+}} dopants (substitutional for \ch{Bi^{3+}}) have an effective $-1$ charge yielding an attractive trapping potential for the positive interstitial \ch{^{8}Li^{+}}, but the \ch{Mn} concentration is an order of magnitude larger.
The high trap density in \gls{btm} will suppress \ch{Li^{+}} mobility.
The exponential increase in $1/T_{1}$ above \SI{200}{\kelvin} may be the onset of a diffusive \gls{bpp} peak, but, in this case, one does not expect it to be so similar between the two very different magnetic fields.
This may reflect a trade-off between the increase in $\omega_{0}$ that shifts the peak to higher temperature, slowing the relaxation on its low-$T$ flank, and the increased polarization of the \ch{Mn} moments by the field that amplifies local magnetic inhomogeneities.
A motional origin for this increase is consistent with the apparent $E_{A} \sim \SI{0.2}{\electronvolt}$, similar to \ch{^{8}Li^{+}} in \gls{bts}~\cite{2019-McFadden-PRB-99-125201}, which also has a \ch{Te} bounded \gls{vdw} gap of similar size to \gls{btm} (\SI{2.620}{\angstrom}).
However, it may have a different explanation, see below in \Cref{sec:discussion:magnetic}.

\subsection{Electronic effects at low temperature \label{sec:discussion:electronic}}

Bismuth chalcogenide (\ch{Bi2Ch3}) crystals exhibit substantial bulk conductivity, despite a narrow gap in the \gls{3d} band structure, making it difficult to distinguish effects of the metallic \gls{tss}.
This is due to native defects (\latin{e.g.}, \ch{Ch} vacancies) that are difficult or impossible to avoid~\cite{2013-Cava-JMCC-1-3176}.
Extrinsic dopants, such as substitutional \ch{Ca/Bi}, can be used to compensate the spontaneous $n$-type doping.
Brahlek \latin{et al.}\ have argued~\cite{2015-Brahlek-SSC-215-54} that, even for the most insulating compensated samples, the carrier densities far exceed the Mott criterion, making them heavily doped semiconductors in the metallic regime.
In this case, we expect metallic \gls{nmr} characteristics~\cite{1978-Holcomb-SUSSP-19-251}, namely a magnetic Knight shift $K$, proportional to the carrier spin susceptibility $\chi_{s}$~\footnote{%
The \gls{nmr} shift $K$ is comprised of several terms including both the spin and orbital response of all the surrounding electrons. In metals, where the local carrier density is substantial,
the Fermi contact coupling with the electron spins often dominates, as we have assumed here.%
}.
In the simplest (isotropic) case,
\begin{equation} \label{eq:knight-shift}
   K = A \chi_{s},
\end{equation}
where $A$ is the hyperfine coupling constant, which is accompanied by a \gls{slr} rate following the Korringa law~\cite{1950-Korringa-P-16-601, 1990-Slichter-PMR},
\begin{equation} \label{eq:korringa-rate}
   \frac{1}{T_{1}} = 4 \pi \hbar A^{2} \gamma_{n}^{2} \left ( \frac{\chi_{s}}{g^{*}\mu_{B}} \right )^{2} k_{B} T.
\end{equation}
Here, $\gamma_{n}$ is the nuclear gyromagnetic ratio, $g^{*}$ is the carrier $g$-factor, and $\mu_{B}$ is the Bohr magneton.
Combining \Cref{eq:knight-shift,eq:korringa-rate}, we obtain the Korringa product, which is independent of the value of $A$,
\begin{equation} \label{eq:korringa-product}
   T_{1}TK^{2} = \frac{\hbar (g^{*} \mu_B)^{2}}{4 \pi k_{B} \gamma_{n}} = S(g^{*}).
\end{equation}
For \ch{^{8}Li},
\begin{equation*}
   S(g^{*}) \approx 1.20 \times 10^{-5} \left ( \frac{ g^{*} }{g_0} \right )^{2}~\si{\second\kelvin},
\end{equation*}
where, unlike in metals, we have allowed for an effective $g$-factor that may be far from its free electron value
$g_0 \approx 2$~\cite{1972-Look-PRB-5-3406}.
Indeed, recent \gls{epr} measurements in \ch{Bi2Se3} find $g^{*} \approx 30$~\cite{2016-Wolos-PRB-93-155114}.

According to Ref.~\onlinecite{2015-Brahlek-SSC-215-54}, \gls{bts} and \gls{bsc} lie on opposite sides of the Ioffe-Regel limit, where the carrier mean free path is equal to its Fermi wavelength, with \gls{bsc} having a higher carrier density and mobility.
A comparative Korringa analysis could test this assertion and, to this end, using \Cref{eq:korringa-product} we define the dimensionless Korringa ratio as
\begin{equation} \label{eq:korringa-constant}
   \mathscr{K} = \frac{T_{1}TK^2}{S(g^{*})}.
\end{equation}
Below the Ioffe-Regel limit, the autocorrelation function of the local hyperfine field at the nucleus, due to the carriers (that determines $T_{1}$) becomes limited by the diffusive transport correlation time.
This has been shown to enhance the Korringa rate (\latin{i.e.}, shortening $T_{1}$)~\cite{1971-Warren-PRB-3-3708, 1983-Gotze-ZPB-54-49}.
From this, one expects $\mathscr{K}$ would be smaller in \gls{bts} than in \gls{bsc}.

There are, however, significant difficulties in determining the experimental $\mathscr{K}$.
First, the Korringa slope depends on magnetic field.
At low fields, this is due to coupling with the host nuclear spins, a phenomenon that is quenched in high fields where the \ch{^{8}Li} \gls{nmr} has no spectral overlap with the \gls{nmr} of host nuclei.
For example, in simple metals, we find the expected field-independent Korringa slope at high fields in the Tesla range~\cite{2015-MacFarlane-SSNMR-68-1}.
In contrast, in \gls{bts}, the slope decreases substantially with increasing field, even at high fields~\cite{2019-McFadden-PRB-99-125201}.
We suggested this could be the result of magnetic carrier freeze-out.
While we do not have comparably extensive data in \gls{bsc}, we can compare the slope at the same field, \SI{6.55}{\tesla} (see \Cref{tab:korringa}).
Here, in both materials, the relaxation is extremely slow, exhibiting no curvature in the \gls{slr} during the \ch{^{8}Li} lifetime
(\Cref{fig:bsc-slr-spectra}), so the uncertainties in $1/T_{1}T$ are likely underestimates.
The larger slope is, however, consistent with a higher carrier density $n$ in \gls{bsc}.
The Korringa slopes should scale~\cite{1972-Look-PRB-5-3406} as $n^{2/3}$.
Using $n \sim \SI{1e19}{\per\centi\meter\cubed}$ in \gls{bsc}~\cite{2009-Hor-PRB-79-195208} and \SI{\sim 2e17}{\per\centi\meter\cubed} in \gls{bts}~\cite{2011-Jia-PRB-84-235206}, the slopes should differ by a factor of \num{\sim 14}, while experimentally the ratio is \num{\sim 5}.

The next difficulty is accurately quantifying the shift $K$, which is quite small with a relatively large demagnetization correction.
Experimentally, the zero of shift, defined by the calibration in \ch{MgO}, differs from the true zero (where $\chi_{s} = 0$) by the difference in chemical (orbital) shifts between \ch{MgO} and the chalcogenide.
However, because \ch{Li} chemical shifts are universally very small, this should not be a large difference, perhaps a few \si{\ppm}.
The negative low temperature shift is also somewhat surprising.
The hyperfine coupling $A$ for \ch{Li} is usually determined by a slight hybridization of the vacant $2s$ orbital with the host conduction band.
As the $s$ orbital has density at the nucleus, the resulting coupling is usually positive, with the $d$ band metals \ch{Pd} and \ch{Pt} being exceptional~\cite{2015-MacFarlane-SSNMR-68-1}.
For a positive $A$, the sign of $K$ is determined by the sign of $g^{*}$, which has not yet been
conclusively measured in either \gls{bsc} or \gls{bts}.
A more serious concern is that $K$ depends on temperature (in contrast to simple metals) and, at least in \gls{bts}, also on applied field~\cite{2019-McFadden-PRB-99-125201}.
To avoid ambiguity from the field dependence, we similarly restrict comparison to the same field, \SI{6.55}{\tesla}.
A similarly temperature dependent shift (for the \ch{^{207}Pb} \gls{nmr}) was found in the narrow band semiconductor \ch{PbTe}~\cite{1973-Hewes-PRB-7-5195}, where it was explained by the temperature dependence of the Fermi level $E_{F}$~\cite{1970-Senturia-PRB-1-4045}.
At low-$T$ in the heavily $p$-type \ch{PbTe}, $E_{F}$ occurs in the valence (or a nearby impurity) band, but with increasing temperature, moves upward into the gap, causing a reduction in $|K|$.
With this in mind, we assume the low temperature shift is the most relevant for a Korringa comparison.
Without a measured $g^{*}$ in \gls{bts}, we simply assume it is the same as \gls{bsc}, and use the $g^{*}_{\parallel}$ from \gls{epr}~\cite{2016-Wolos-PRB-93-155114} to calculate the values of $\mathscr{K}$ in \Cref{tab:korringa}.

\begin{table}
\centering
\caption[
Korringa analysis of \acrlong{bsc} and \acrlong{bts} at \SI{6.55}{\tesla} and low temperature.
]{ \label{tab:korringa}
Korringa analysis of \gls{bsc} and \gls{bts}~\cite{2019-McFadden-PRB-99-125201} at \SI{6.55}{\tesla} and low temperature.
To calculate $\mathscr{K}$, we take $S(g^{*})$ in \Cref{eq:korringa-constant} to be \SI{2.69e-3}{\second\kelvin}, using the $g_{\parallel}^{*}$ from \gls{epr}~\cite{2016-Wolos-PRB-93-155114}.
}
\begin{tabular}{l S S S}
\toprule
    & {$1/(T_{1}T)$ (\SI{e-6}{\per\second\per\kelvin})} & {$K$ (\si{\ppm})} & {$\mathscr{K}$} \\
\midrule
\acrlong{bsc} & 9.5  \pm 0.8  &  -31 \pm 3 &   0.038 \pm 0.008 \\
\acrlong{bts} & 1.79 \pm 0.07 & -115 \pm 3 &   2.78  \pm 0.18  \\
\bottomrule
\end{tabular}
\end{table}

The values of $\mathscr{K}$ are just opposite to the expectation of faster relaxation for diffusive \gls{bts} compared to metallic
\gls{bsc}~\cite{2015-Brahlek-SSC-215-54}.
Electronic correlations can, however, significantly alter the Korringa ratio to an extent that depends on disorder~\cite{1994-Shastry-PRL-72-1933}.
There should be no significant correlations in the broad bulk bands of the chalcogenides, but in narrow impurity bands, this is certainly a possibility.
We note that $\mathscr{K}$ is also less than \num{1} in \ch{PbTe}~\cite{1973-Alexander-JN-1-251}, similar to \gls{bsc}.
At this stage, without more data and a better understanding of the considerations mentioned above, it is premature to draw further conclusions.

\subsection{Magnetism in \acrlong{btm} \label{sec:discussion:magnetic}}

In the \ch{Mn} doped \ch{Bi2Te3} at low field, the relaxation from magnetic \ch{Mn^{2+}} becomes faster as the spin fluctuations slow down on cooling towards $T_{C}$.
In particular, the low-$T$ increase in $1/T_{1}$ occurs near where correlations among the \ch{Mn} spins become evident in \gls{epr}~\cite{2016-Zimmermann-PRB-94-125205, 2017-Talanov-AMR-48-143}.
In remarkable contrast to ferromagnetic \ch{EuO}~\cite{2015-MacFarlane-PRB-92-064409}, the signal is not wiped out in the vicinity of $T_{C}$, but $1/T_{1}$ does become very fast.
This is likely a consequence of a relatively small hyperfine coupling consistent with a \ch{Li} site in the \gls{vdw} gap.

High applied field slows the \ch{Mn} spins more continuously starting from a higher temperature, suppressing the critical peak and reducing $1/T_{1}$ significantly, a well-known phenomenon in \gls{nmr} and \gls{musr} at magnetic transitions (see \latin{e.g.}, Ref.~\onlinecite{2001-Heffner-PRB-63-094408}).
This also explains the small critical peak in the \ch{^{8}Li} \gls{slr} in the dilute magnetic semiconductor, \ch{Ga_{1-x}Mn_{x}As}~\cite{2011-Song-PRB-84-054414}.
As in \ch{GaAs}, \ch{Mn} in \ch{Bi2Te3} is both a magnetic and electronic dopant.
At this concentration, \gls{btm} is $p$-type with a metallic carrier density of \SI{\sim 7e19}{\per\centi\meter\cubed}~\cite{2010-Hor-PRB-81-195203, 2016-Zimmermann-PRB-94-125205}.
However, the difference in scale of $1/T_{1}$ at high field between \Cref{fig:btm-slr-fits,fig:bsc-slr-fits} shows that the \ch{Mn} spins completely dominate the carrier relaxation.

It is also remarkable that the resonance is so clear in the magnetic state, in contrast to \ch{Ga_{1-x}Mn_{x}As}~\cite{2011-Song-PRB-84-054414}.
The difference is not the linewidth, but rather the enhanced amplitude.
This may be due to slow \ch{^{8}Li} spectral dynamics occurring on the timescale of $\tau_{\beta}$ that effectively enhance the amplitude, 
for example, slow fluctuations of the ordered \ch{Mn} moments, not far below $T_{C}$.
Similar behavior was found in rutile \ch{TiO2} at low temperature, where it was attributed to field fluctuations due to a nearby electron polaron~\cite{2017-McFadden-CM-29-10187}.
Enhancement of the \gls{rf} field at nuclei in a ferromagnet~\cite{1959-Gossard-PRL-3-164} may also play a role.

Above \SI{200}{\kelvin}, the activated increase in $1/T_{1}$ may indicate the onset of diffusive \ch{^{8}Li^{+}} motion, similar to the nonmagnetic analogs, with the additional effect that the local magnetic field from the \ch{Mn} spins is also modulated by \ch{^{8}Li} hopping (not just the \gls{efg}), similar to the ordered magnetic \gls{vdw} layered \ch{CrSe2}~\cite{2020-Ticknor-RSCA-10-8190}.
However, it may instead mark the onset of thermal excitation of electrons across the bandgap that is narrowed by \ch{Mn} doping\cite{2010-Hor-PRB-81-195203}.
Thermally excited carriers may also explain the increase in $1/T_1$ at comparable temperatures in \ch{Ga_{1-x}Mn_{x}As}~\cite{2011-Song-PRB-84-054414}, a very different medium for \ch{Li^+} diffusion.
Thermally increased carrier density would strengthen interaction between the \ch{Mn} moments, extend their effects via \gls{rkky} polarization, and increase $1/T_{1}$.
Measurements at higher temperatures may be able to discriminate these possibilities, but it is likely that both processes will contribute.

The stretched \gls{slr} and field-dependent $1/T_{1}$
are characteristic of the \gls{nmr} response in a disordered glassy magnetic state, consistent with
the random Mn/Bi site disorder. The magnetic properties of \gls{btm}~\cite{2010-Hor-PRB-81-195203, 2013-Watson-NJP-15-103016, 2016-Zimmermann-PRB-94-125205, 2019-Vaknin-PRB-99-220404} are similar to the dilute (ferro)magnetic semiconductors that include a uniform magnetization of the carriers. In this context, our data provides a \emph{local} characterization of the inhomogeneous magnetic state of \gls{btm} that will be useful in developing a detailed microscopic understanding of its magnetism.
Having established the effects of \ch{Mn} magnetism in the bulk, it would be interesting to use lower implantation energies to study how they may be altered in the surface region by coupling to the \gls{tss}.

\section{Conclusion \label{sec:conclusion}}

Using implanted \ch{^{8}Li} \gls{bnmr}, we have studied the electronic and magnetic properties of the doped \glspl{ti} \gls{bsc} and \gls{btm}.
From \gls{slr} measurements, we find evidence at temperatures above \SI{\sim 200}{\kelvin} for site-to-site hopping of isolated \ch{Li^{+}} with an Arrhenius activation energy of \SI{\sim 0.4}{\electronvolt} in \gls{bsc}.
At lower temperature the electronic properties dominate, giving rise to Korringa-like relaxation and negative Knight shifts, similar to isostructural \gls{bts}.
A quantitative comparison reveals Korringa ratios opposite to expectations across the Ioffe-Regel limit.
In \gls{btm}, the magnetism from dilute \ch{Mn} moments dominates all other spin interactions, but the \gls{bnmr} signal remains measurable through the magnetic transition at $T_{C}$, where a critical peak in the \gls{slr} rate is observed.
The \SI{\sim 0.2}{\electronvolt} activation energy from the high temperature increase in the \gls{slr} may be related to \ch{Li} mobility or to thermal carrier excitations.

With these new results, a more complete picture of the implanted \ch{^{8}Li} \gls{nmr} probe of the tetradymite \ch{Bi} chalcogenides (and other \gls{vdw} chalcogenides) is beginning to emerge.
At high temperatures, isolated \ch{^{8}Li^{+}} has a tendency to mobilize, providing unique access to the kinetic parameters governing \ch{Li^{+}} diffusion in the ultra-dilute limit.
At low temperature, \ch{^{8}Li} is sensitive to the local metallic and magnetic properties of the host.
With the bulk \gls{nmr} response now established in \ch{Bi2Ch3} \glspl{ti}, the prospect of directly probing the chiral \gls{tss} with the depth resolution provided by \gls{bnmr} remains promising.

\begin{acknowledgments}
We thank:
R.\ Abasalti, D.\ J.\ Arseneau, S.\ Daviel, B.\ Hitti, K.\ Olchanski, and D.\ Vyas for their excellent technical support;
M.\ H.\ Dehn, T.\ J.\ Parolin, O.\ Ofer, Z.\ Salman, Q.\ Song, and D.\ Wang for assistance with early measurements;
as well as D.\ E.\ MacLaughlin, S.\ D.\ Senturia, and A.\ Wolos for useful discussions.
This work was supported by NSERC Discovery grants to R.F.K.\ and W.A.M.
Additionally, R.M.L.M.\ and A.C.\ acknowledge support from their NSERC CREATE IsoSiM Fellowships.
The crystal growth at Princeton University was supported by the ARO-sponsored MURI on topological insulators, grant number W911NF1210461.
\end{acknowledgments}

\appendix

\section{\ch{^{8}Li^{+}} implantation profiles \label{sec:implantation}}

\begin{figure}
\centering
\includegraphics[width=\columnwidth]{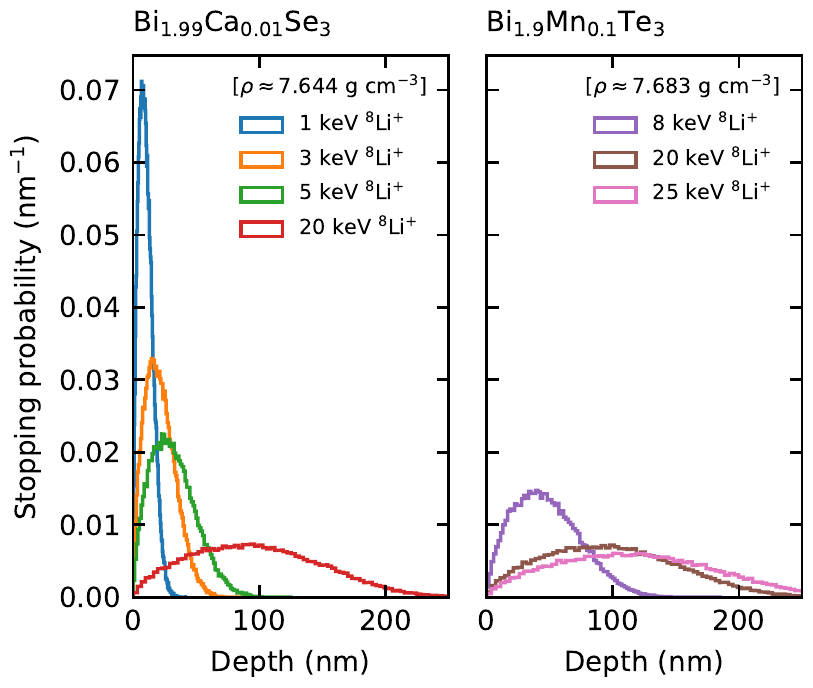}
\caption[
Stopping distribution for \ch{^{8}Li^{+}} implanted in the doped \acrlongpl{ti} \acrlong{bsc} and \acrlong{btm}.
]{ \label{fig:dti-srim}
Stopping distribution for \ch{^{8}Li^{+}} implanted in the doped \glspl{ti} \gls{bsc} and \gls{btm}, calculated using the \gls{srim} Monte Carlo code~\cite{srim} using \num{e5} ions.
The implanted probe ions stop principally within \SI{\sim 250}{\nano\meter} from the crystal surface, with only a minor ``tail'' penetrating further.
Only at the lowest \ch{^{8}Li^{+}} implantation energy \SI{1}{\kilo\electronvolt} does an appreciable fraction (\SI{\sim 15}{\percent}) stop in the region where surface effects,
are expected to be important.
}
\end{figure}

\begin{figure}
\centering
\includegraphics[width=\columnwidth]{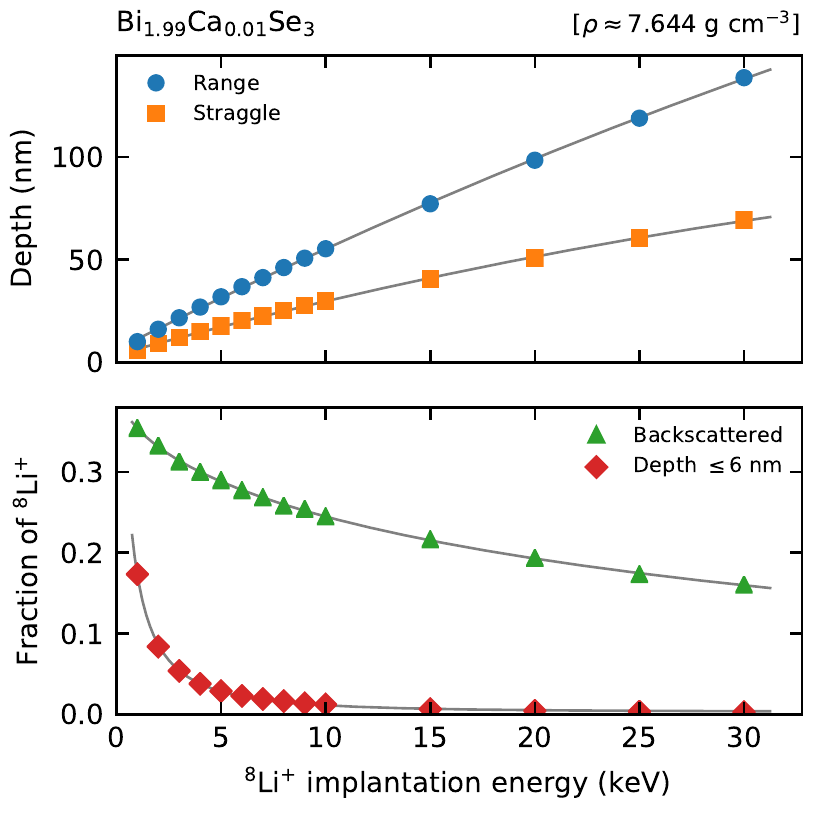}
\caption[
Stopping details for \ch{^{8}Li^{+}} implanted in \acrlong{bsc}.
]{ \label{fig:bsc-srim-range}
Stopping details for \ch{^{8}Li^{+}} implanted in \gls{bsc} calculated using the \gls{srim} Monte Carlo code~\cite{srim}.
The ion stopping range and straggle (\latin{i.e.}, mean and standard deviation) at each implantation energy are shown in the top panel.
The bottom panel shows the predicted fraction of \ch{^{8}Li^{+}} that are backscattered and those that stop within the first \SI{\sim 6}{\nano\meter}, where surface effects are expected.
Except at the lowest implantation energy, this latter fraction is negligible.
The solid grey lines are drawn to guide the eye.
}
\end{figure}

The \gls{srim} Monte Carlo code~\cite{srim} was used to predict the \ch{^{8}Li^{+}} implantation profiles in \gls{bsc} and \gls{btm}.
At a given implantation energy, stopping events were simulated for \num{e5} ions, which were histogrammed to represent the predicted implantation profile shown in \Cref{fig:dti-srim}.
From the stopping profiles, we calculated, in the nomenclature of ion-implantation literature, the range and straggle (\latin{i.e.}, the mean and standard deviation) of the \ch{^{8}Li^{+}} stopping depth.
Additionally, the fraction of the simulated ions that were backscattered (\latin{i.e.}, did not stop in the target material) or stopped at very shallow depths below the surface, were determined for \gls{bsc} (see \Cref{fig:bsc-srim-range}).
At the ion beam energies used here (\SIrange{1}{25}{\kilo\electronvolt}), at most a small minority of the implanted \ch{^{8}Li^{+}} ions stop in the top few \si{\nano\meter} where surface effects are anticipated.
Thus, we do not expect a significant contribution from the \gls{tss}. To study it will require an implantation energy substantially less than \SI{1}{\kilo\electronvolt}.

\section{Helicity-resolved resonances \label{sec:helicities}}

As mentioned in \Cref{sec:experiment}, unique to ion-implanted \gls{bnmr},
the probe nuclear spin polarization is produced \latin{ex situ}
using collinear optical pumping~\cite{2014-Levy-HI-225-165}.
The projection of the nuclear spin on the beam direction (the helicity) is either ``positive'' or ``negative'' depending on the sense (left or right) of the circularly polarized laser light,
which is alternated during each measurement.
Data are collected separately for each laser helicity yielding helicity-resolved spectra,
which are useful at elucidating details of the resonance~\cite{2015-MacFarlane-SSNMR-68-1}.

\begin{figure}
\centering
\includegraphics[width=\columnwidth]{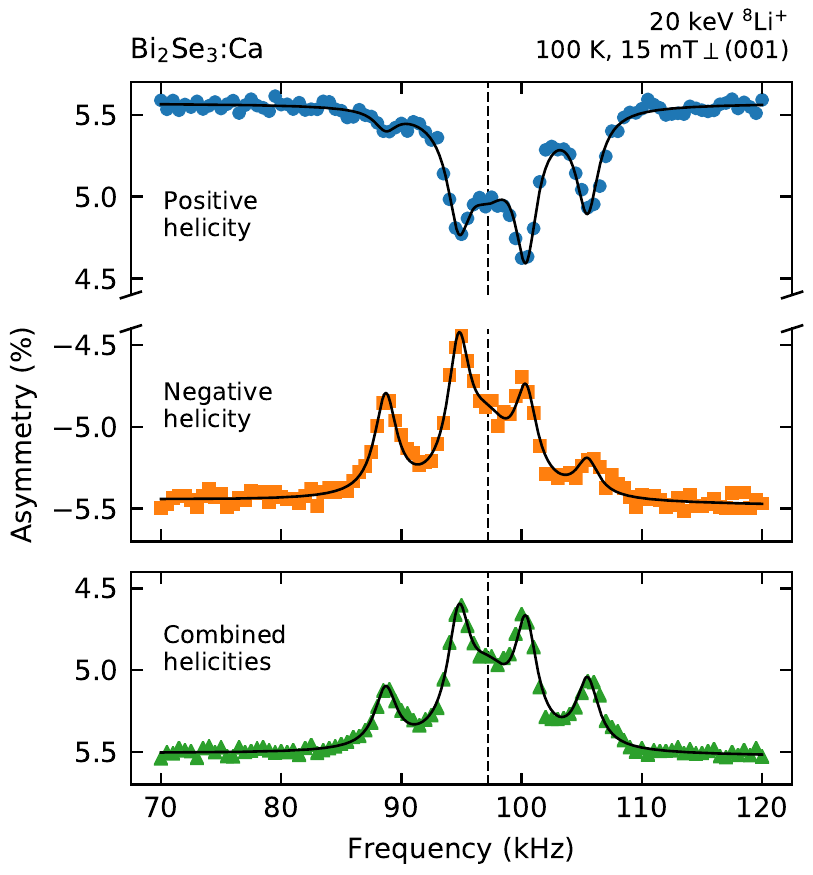}
\caption[
Typical helicity-resolved \ch{^{8}Li} spectra in \acrlong{bsc}, revealing the fine structure of the line.
]{ \label{fig:bsc-1f-spectra-helicities}
Typical helicity-resolved \ch{^{8}Li} spectra in \gls{bsc}, revealing the fine structure of the line.
Four quadrupole satellites, split asymmetrically in each helicity about a ``central'' Lorentzian (marked by a vertical dashed line) are evident.
Note that, in contrast to conventional \gls{nmr}, the satellite amplitudes are determined primarily by the highly polarized initial state.
The solid black lines are the fits described in the text and the vertical dashed line marks the resonance central frequency $\nu_{0}$.
}
\end{figure}

A typical helicity-resolved \ch{^{8}Li} resonance spectrum in \gls{bsc} is shown in \Cref{fig:bsc-1f-spectra-helicities}.
Besides the well-resolved fine structure, the spectra unambiguously reveal the multi-component nature of the line.
A quadrupolar splitting, giving the anti-symmetric shape about the resonance centre-of-mass in each helicity, on the order of several \si{\kilo\hertz}, can be identified from the outermost satellites.
In contrast to conventional \gls{nmr}, note that the satellite intensities are chiefly determined by the high initial polarization, causing an increase in the relative amplitude of the outer satellites~\cite{2014-MacFarlane-JPCS-551-012059}, which also depend on the \gls{slr}.
Apart from the quadrupolar component, another contribution at the resonance ``centre'', with no resolved splitting, is discernible.
This is consistent with the \ch{^{8}Li} resonances observed in other \gls{vdw} materials~\cite{2006-Wang-PB-374-239, 2019-McFadden-PRB-99-125201, 2020-Ticknor-RSCA-10-8190}.
Unlike the more common case encountered for the spin $I = 3/2$ \ch{^{7}Li} nucleus, there is no $m_{\pm 1/2} \leftrightarrow m_{\mp 1/2}$ main line transition that is unshifted in first-order by the quadrupolar interaction.
The absence of any peaks interlacing the $m_{\pm 2} \leftrightarrow m_{\pm 1}$ transitions suggest this ``central'' line cannot be multi-quantum (\latin{cf.}\ the spectrum in \ch{Bi}~\cite{2014-MacFarlane-JPCS-551-012059, 2014-MacFarlane-PRB-90-214422}).
Instead, the ``central'' peak must originate from \ch{^{8}Li} in an environment where
the static (i.e., time-average) \gls{efg} is close to zero.
For presentation, we combined the helicity-resolved spectra to give an overall average lineshape (\Cref{fig:bsc-1f-spectra-lf} in \Cref{sec:results:bsc}).

\end{document}